\documentclass[10pt, a4paper]{article}
\usepackage{authblk}
\usepackage{amssymb}
\usepackage{amsmath}
\usepackage{latexsym}
\usepackage{tikz}
\usepackage{sidecap,epstopdf}
\usepackage{subfigure}
\usepackage{color}
\usepackage{enumerate}
\usepackage{appendix}
\usepackage{amsthm}\theoremstyle{remark}
\usepackage{hyperref}
\hypersetup{
    colorlinks=true,
    linkcolor=blue,
    citecolor = magenta,
    }

\newcommand{\bte}{\begin{quote}\begin{theorem}}
\newcommand{\ete}[1]{\label{#1}\end{theorem}\end{quote}}
\newcommand{\bcom}{\begin{quote}\begin{comment}}
\newcommand{\ecom}[1]{\label{#1}\end{comment}\end{quote}}
\newcommand{\bex}{\begin{quote}\begin{example}}
\newcommand{\eex}[1]{\label{#1}\end{example}\end{quote}}
\newcommand{\bcon}{\begin{quote}\begin{conclusion}}
\newcommand{\econ}[1]{\label{#1}\end{conclusion}\end{quote}}
\newcommand{\bdefi}{\begin{quote}\begin{definition}}
\newcommand{\edefi}[1]{\label{#1}\end{definition}\end{quote}}

\newcommand{\blem}{\begin{quote}\begin{lemma}}
\newcommand{\elem}[1]{\label{#1}\end{lemma}\end{quote}}

\newcommand{\bpr}{\begin{quote}\begin{problem}}
\newcommand{\epr}[1]{\label{#1}\end{problem}\end{quote}}

\newcommand{\beq}{\begin{eqnarray}}
\newcommand{\eeq}[1]{\label{#1}\end{eqnarray}}

\newcommand{\bfi}{\begin{figure}[24]}
\newcommand{\efi}[1]{\caption{\label{#1}}\end{figure}}

\newtheorem{thm}{Theorem}
\theoremstyle{remark}
\newtheorem{rem}[thm]{Remark}
\newtheorem{ex}[thm]{Example}

\newcommand{\ptl}{\partial}

\newcommand{\rmU}{{\rm U}}
\newcommand{\rmV}{{\rm V}}
\newcommand{\rmA}{{\rm K}}
\newcommand{\rmF}{{\rm F}}
\newcommand{\rmX}{{\rm X}}
\newcommand{\rmE}{{\rm E}}
\newcommand{\rmC}{{\rm C}}
\newcommand{\kstar}{z_i}

\newcommand{\bexe}{\begin{quote}\begin{exercise}\inh}
\newcommand{\eexe}[1]{\label{#1}\end{exercise}\end{quote}}

\graphicspath{{./figures/}}
\usepackage{graphics,graphicx}
\usepackage{color}

\author[2]{A. I. Korolkov, A. V. Kisil}

\title{Recycling solutions of boundary value problems: 
the Wiener--Hopf perspective on embedding formula}

\date{}
\begin{document}
\maketitle
\begin{abstract}
Embedding formula allows to recycle solution of a family boundary value problems by expressing all the solutions in terms of a small number of solutions. Such formulas have been previously derived in the context of diffraction by applying a cleverly chosen operator to the solution and the construction of edge Green's functions which are introduced in an elaborate manner specific for each problem. We demonstrate that embedding formula naturally appears from a matrix Wiener--Hopf equation, and the embedding formula is derived from the canonical solution to this matrix Wiener--Hopf problem. This allows to drive the embedding formula in any context where the problem can be formulated as a Wiener--Hopf equation. We illustrate the effectiveness of this approach by revisiting known problems, such as the problem of diffraction by half-line, a strip and the problem of diffraction by a wedge. Additionally, a new matrix Wiener--Hopf formulation is derived for wedge problems.
\end{abstract}

\section{Introduction}

Due to the importance of the boundary value problems (BVPs) there have been a wealth of techniques developed to solve them. Some methods are analytical methods which express the solution in terms of some standard functions and some are numerical methods which produce values of the solution at specific points. One advantage of analytical over numerical solutions of BVPs is that in the former the dependence on the parameters of the BVP is explicit. This is useful for  finding optimal regimes and to address some inverse problems. Of course, there are many methods which are neither fully analytical nor numerical. In this paper, we will consider an intermediate method, called embedding, where the full analytic solutions are not found but it is still possible to obtain some analytical dependence of the solution on some parameter. Specifically, this applies BVPs which can be formulated in terms of Wiener--Hopf equation~\cite{NSR_20}. The Wiener--Hopf equation arises as a generalisation of the Fourier transform based methods and has been extensively used for boundary value problems speciously when boundary conditions have edges and corners. These equations can be derived routinely but their analytical solution is not always possible. Some research areas where it has been used are acoustics, finance, L\'{e}vi processes, hydrodynamics, elasticity, potential theory and electromagnetism. Here embedding method is illustrated by considering diffraction problems with the parameter being the incidence angle of a plane wave.

For example, consider a diffraction problem governed by the Helmholtz equation with given boundary conditions on the obstacles and radiation conditions at infinity. It is forced by an incidence plane wave at an angle \(\theta_i\) (for example see Figure~\ref{fig:strip}). Suppose we would like to calculate the solution for different \(\theta_i\), would we need to solve each for each \(\theta_i\) afresh? For a standard numerical solver such as COMSOL, we would need to re-run it for each of the \(\theta_i\). Parts of the calculations can be reused for a narrow class of problems for which an analytic solution exists. For a wider class of diffraction problems, there is a link between the solutions for different \(\theta_i\) provided by the so-called embedding formulas. More specifically, the embedding formula allows the solution for a given \(\theta_j\) to be expressed as essentially a linear combination of $N$ solutions with fixed $\theta_i$~\cite{Craster2003}. The coefficients in the linear combination can be determined for the far-field of the solution. Deriving embedding formulas for different BVPs will be the focus of this paper.

\begin{figure}[ht]
	\centering{
	\includegraphics[width=1\textwidth]{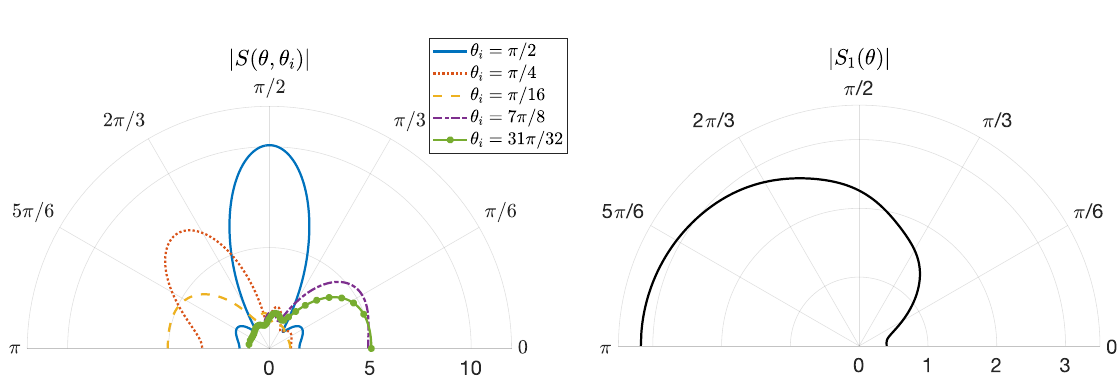}
	}
	\caption{Directivities for the strip problem, with $ka=10$ (left: for the plane wave incidence problem, right: for the edge Green's function).}
	\label{fig:directivity}
\end{figure}

\begin{ex}
We will illustrate the concept of embedding on a well-known problem of diffraction by a finite strip.   The geometry of the problem is shown in Figure~\ref{fig:strip}. 
\begin{figure}[ht]
	\centering{
	\includegraphics[width=0.3\textwidth]{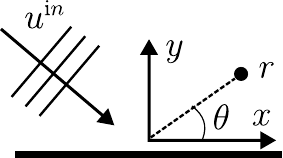}
	}
	\caption{Geometry of the problem of diffraction of a plane wave incident at angle $\theta_i$ by a finite strip.}
	\label{fig:strip}
\end{figure}
Analytical solution can be written in terms of an infinite expansion of special functions called Mathieu functions~\cite{Sieger1908, Colbrook20}. But a natural modification of the problem with more than one co-linear strip has no analytical solution~\cite{iter_n,Shanin2003}. Nevertheless,  a simple modification allows us to obtain the embedding formula for any number of strips~\cite{Shanin2003} showing the adaptability of the method. 

Let the total field $u$ satisfy the Helmholtz equation 
\begin{equation}
\label{eq:Helm}
(\Delta +k^2)u(x,y) = 0,
\end{equation}
where $k$ is a wavenumber. Let the total field satisfy Dirichlet boundary conditions on the strip
\[
u(x,0) = 0, \quad x \in [-a,a].
\]
The problem should also be supported with radiation condition \cite{tikhonov1948radiation} and Meixner tip conditions \cite{Nazarov1994} to possess a unique solution (see \cite{Babic2008}).

Let there be an incident plane wave at an angle \(\theta_i\) 
\[u^{\rm in}(x,y) = \exp\{-ikx\cos\theta_i - iky\sin\theta_i\}.\]
The total field can be presented as a sum of the incident wave $u^{\rm in}$ and the scattered wave $u^{\rm sc}$,
\begin{equation}
\label{eq:inc_field}
u(x,y) = u^{\rm in}(x,y) + u^{\rm sc}(x,y).
\end{equation}
We are interested in finding \(u^{\rm sc}(x,y)\) especially far from the strip.
Introduce polar coordinates
\[
x = r\cos\theta,\quad y= r\sin\theta.
\]
The directivity of the scattered field denoted by \(S(\theta,\theta_i)\) is given by
\begin{equation}
\label{eq:Direct_sc}
u^{\rm sc}(r,\theta) = -\frac{\exp\{ikr-i\pi/4\}}{\sqrt{2\pi k r}}S(\theta,\theta_i)+O\left(\frac{1}{r}\right).
\end{equation}
Note that the directivity of the scattered field does not depend on \(r\) and provides the large \(r\) behaviour which is important for many applications. 

Then, let us introduce auxiliary problems for the edge Green's functions. Particularly, instead of plane wave forcing let us consider two problems with point sources placed in the vicinity (to be more precise a limit as the source moves to the edge is considered) of the left and right edge of the strip correspondingly (see~Figure~\ref{fig:edge_Green}).
\begin{figure}[ht]
	\centering{
	\includegraphics[width=0.7\textwidth]{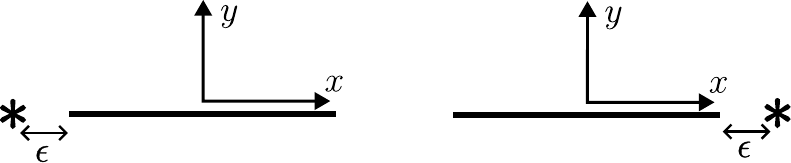}
	}
	\caption{Geometry of problems for edge Green's functions. There is a point source located either on the left or the right of the strip.}
	\label{fig:edge_Green}
\end{figure}
Denote the directivities of edge Green's functions as $S_1(\theta)$ and $S_2(\theta)$ correspondingly. Then, the embedding formula is given by the following relation \cite{Craster2003}:
\begin{equation}
\label{eq:Embed_strip_dir}
S(\theta,\theta_i) = \frac{S_2(\theta_i)S_2(\theta) - S_1(\theta_i)S_1(\theta)}{k\cos\theta + k\cos\theta_i}.
\end{equation}
The main computational feature of the latter is that the left-hand side is a function of two parameters and the right-hand side involves only functions of one parameter. Note that $S_1(\theta)$ and $S_2(\theta)$ are just reflections of each other due to the symmetry of the problem, this also allows to avoid the numerical problem of zeros in the denominator which can occur in other embedding formulas~\cite{Gibbs_2018}. So computing $S_1(\theta)$ allows to construct easily all the \(S(\theta,\theta_i)\) via the formula, some are plotted in Figure~\ref{fig:directivity}.
We will give a more formal definition for the edge Green's function (Section~\ref{sec:Edge Green}), and show how (\ref{eq:Embed_strip_dir}) can be derived using the Wiener--Hopf method in Section~\ref{sec:strip}.
\end{ex}
There are two main forms of embedding formula one we will call canonical which expresses the solution in terms of edge Green's functions  (oversingular solutions) and the second one which does so in terms of plane waves, which we will call plane wave embedding. 
Papers on embedding in diffraction theory usually quote 1982 paper by  M. H. Williams as the foundational work \cite{WILLIAMS1982}. Early developments by Porter et al. \cite{porter1991solution} were based on integral equations formulation and operator manipulations. 
Now the standard way to derive the embedding formula is via the application of an operator to the governing problem, to derive canonical embedding see the steps below~\cite{Craster2003}:
 \begin{enumerate}
 \item Find operator, \(H\), so that applied to the total field \(u\), the result \(H [u]\) is still a solution to the Helmholtz equation \eqref{eq:Helm} with the same boundary conditions (excluding the tip conditions) and \(H [u^{\rm in}]=0\), where \(u^{\rm in}\) is the incident plane wave. 
 \item Define the correct edge Green's functions \(v_i\) (oversingular solutions) by placing line source(s) near the tips so that they have the same behaviour as \(H [u]\) at the tips.
 \item Show that \(L[u,v_i]=H [u]+\sum_i K_i v_i\) satisfies the tip conditions, with some constants  $K_i$ that are to be determined. Since  \(L[u,v_i]\) is the solution of the diffraction problem, then, by uniqueness \(L[u,v_i]=0\), giving the \emph{weak form of embedding}.
 \item  Take the far-field limit and use reciprocity to obtain the \emph{strong embedding formula}.
 \end{enumerate}
In~\cite{Biggs2006} a plane wave embedding was proposed by modifying Step 2 of the procedure above. Specifically, instead of the edge Green's functions, the author used the plane wave solutions  $\tilde u_i$ with different incidence angles and showed that  \(H [u]+\sum_i \tilde K_i\hat{u}_i=0\) still held in this case.

 Here we propose a different way of deriving the embedding formula via the Wiener--Hopf formulation~\cite{Noble1958, myWHreview}. The advantage of this approach is that the derivations are simple algebraic manipulations once the Wiener--Hopf equations are obtained. They also offer a way of obtaining some information about the solution in the case where the full Wiener--Hopf system cannot be solved. ions of the embedding formula, there is no need to perform factorisation of the kernel which is typic --- the only step in the Wiener--Hopf procedure that generally cannot be performed analytically. It is also worth noting that the edge Green's function/over singular solutions that appear in the standard derivations are linked to the normal solutions~\cite{gakhov1952riemann, Muskhelishvili1977} of the homogeneous Wiener--Hopf problem. Normal solutions have also been used in the context of Fredholm factorization~\cite{Dan_Lomb07}. Pole removal then allows us to obtain a solution of the full Wiener--Hopf system in terms of the normal solutions. Note that almost all cases in which the embedding is known are also the cases where the problems can be formulated as a Wiener--Hopf system. Lastly, although this embedding method originated in acoustics via the Wiener--Hopf formulation it can now be used in electromagnetism, Lévi processes and potential theory.

The paper is organised as follows. The formulation of the inhomogeneous matrix Wiener--Hopf problem and auxiliary homogeneous problems (normal solutions) are given in Section~\ref{sec: prelim}. General way of deriving the embedding formula via the Wiener--Hopf equation is given in Section~\ref{sec: can_embed}. The new approach is demonstrated in Section~\ref{sec: examples} on well-studied problems of diffraction theory: the half-plane, the finite strip, and the wedge problems. This also motivated a new matrix formulation for the wedge problem Section~\ref{sec: wedge}. A more generalized way of deriving of embedding formula for Wiener-Hopf equations that involve additional change of variable is presented in Section~\ref{map}. The approach is demonstrated on the problem of diffraction by a wedge.  Discussions and comparisons are placed in the last section.

\section{Preliminaries}
\label{sec: prelim}
\subsection{Problem formulation}
We study the following inhomogeneous matrix Wiener--Hopf problem:
\begin{equation}
\label{eq: matrix_WH}
\rmU^-(t,z_i)  = {\rmA}(t) {\rmU}^+(t,z_i) + \rmF(t,z_i),\quad t \in     \mathbb{R},
\end{equation}
where ${\rmU}^-(z,z_i)$ /${\rmU}^+(z,z_i)$ is a vector function of size $N$ analytic in the lower/upper half-plane and having at most polynomial growth at infinity, ${\rmA}(z)$ is known matrix function of size $N\times N$ that referred as the kernel of the Wiener--Hopf equation, ${\rmF}(t,z_i)$ is a vector of size $N$ of forcing terms, and $z_i$ is a complex parameter. Specifically, we are interested in the particular form of forcing
\begin{equation}
  \label{eq:pole_forcing}
{\rmF}(z,z_i) = \frac{{\rm r}}{z-z_i},\quad {\rm Im}[z_i]<0,
\end{equation}
where ${\rm r}$ is some constant vector. In the context of diffraction theory, (\ref{eq:pole_forcing}) corresponds to forcing by a plane wave incidence, and $z_i$ is associated with the angle of incidence. 
% $\theta_i$ as follows: 
% \begin{eqnarray}
% z_i = k\cos \theta^{\rm i}.    
% \end{eqnarray}

\subsection{Normal solutions to the Wiener--Hopf equations}

In this section, we define a system of normal solutions of (\ref{eq: matrix_WH}), which will be used later to obtain the canonical embedding formula. Using the problem of diffraction by a finite strip as an example, we show that edge Green's functions can be interpreted as normal solutions for the strip problem.  

In 1950s Muskhelishvili and Gakhov developed a method of solution \cite{gakhov1952riemann, Muskhelishvili1977} of the matrix Riemann-Hilbert problem for an arbitrary forcing. The key subject that they introduced was the system of normal solutions. Below we adapt their definitions for the problem (\ref{eq: matrix_WH}), and show that for the particular case of forcing (\ref{eq:pole_forcing}) their approach gives rise to the embedding formula. 
% Find column vectors of size \(N\), ${\rm V}^{\pm}(z)$ analytic in upper/lower half-planes that satisfy the following  equation:

Consider a homogeneous Wiener--Hopf problem, i.e (\ref{eq: matrix_WH}) with $\rmF={\rm 0}$:
\begin{equation}
\label{eq: Wh_homo}
{\rmV}^-_{(j)}(t)={\rmA}(t){\rmV}^+_{(j)}(t),\quad t \in \mathbb{R}, \quad j = 1,\ldots,N.
\end{equation}
Note that solutions $V^\pm_{(j)}$ do not depend on parameter $z_i$.  A system of $N$ solutions to the above  $\rmV^{\pm}_{(n)}(z)$, \(n=1, \dots, N\) is called \textit{normal}  if matrix
\begin{equation*}
{\rmX}^{\pm}(z) = ({\rmV}^{\pm}_{(1)}(z),\ldots,{\rmV}^{\pm}_{(N)}(z)),
\end{equation*}
do not have zeroes of its determinant on the whole complex plane (except possibly at infinity). Matrix ${\rmX}$ is referred as normal.

Normal solutions provide a way to express an arbitrary solution of the Wiener-Hopf equation. Given a set of normal solutions  ${\rm V^{\pm}}_{(n)}(z)$, \(n=1, \dots, N,\)
  all the other solutions of the homogeneous matrix Wiener--Hopf problem can be written as
  \[\sum_{i=1}^{N} P_i(z){\rm V^{\pm}}_{(i)}(z),\]
  where \( P_i(z)\) are polynomials. %\\[0.2cm]
The key observation that we will need is that normal solutions provide a factorisation of the kernel ${\rmA}(z)$. Hence solutions of the inhomogeneous equation can also be expressed in terms of normal solutions, which we use in Section~\ref{sec: can_embed}.

\begin{rem}
In the examples below we will also impose growth conditions along \eqref{eq: Wh_homo} this will ensure that the solution is unique and that it corresponds to the physically motivated edge Green's functions described in the next section. 
\end{rem}

\subsection{Edge Green's functions as normal solutions}
\label{sec:Edge Green}

Let us show that for the problem of diffraction by a finite strip, discussed in the introduction, edge Green's functions give rise to a normal system of solutions for the associated Wiener--Hopf problem. Indeed, standard calculations show (see~Appendix~\ref{app: strip}) that the problem of diffraction by a finite strip can be reduced to matrix Wiener--Hopf equation (\ref{eq: matrix_WH})
with
\[
{\rm U^\pm}=
\begin{pmatrix}
U^\pm_1 \\
U^\pm_2
\end{pmatrix},
\quad
{\rm A}=
\begin{pmatrix}
-\exp\{2iz a\} &  (i\gamma(z))^{-1} \
0 &\exp\{-2iz a\}
\end{pmatrix},\quad
{\rm F} = \frac{i \exp\{i\kstar a\}}{z-\kstar}
\begin{pmatrix}
1\\
0
\end{pmatrix},
\]
\[
U^-_1(z,z_i)  = \exp\{iz a\}\int_{-\infty}^{-a} \exp\{iz x\}u^{\rm sc}(x,0) dx,
\]
\[
U^-_2(z,z_i)  = \exp\{iz a\}\int_{-a}^{a} \exp\{iz x\}\frac{\ptl u^{\rm sc}}{\ptl y}(x,0) dx,
\]
\[
U^+_1(z,z_i) = \exp\{-iz a\}\int_a^\infty \exp\{iz x\}u^{\rm sc}(x,0)dx + \frac{i\exp\{-i\kstar a\}}{z-\kstar}, 
\]
\[
U^+_2(z,z_i)  = \exp\{-iz a\}\int_{-a}^{a} \exp\{iz x\}\frac{\ptl u^{\rm sc}}{\ptl y}(x,0) dx,
\]
and $\gamma(z) = \sqrt{k^2 - z^2}$.
The latter should be supported by the growth conditions which follow from Meixner's edge conditions via Watson's lemma \cite{Olver2014}:
\[
U^{-}_1\sim {\frac{C_1}{z^{3/2}}}, \quad
U^{-}_2 \sim \frac{C_2}{\sqrt{z}}, \quad 
U^{+}_1 \sim {\frac{i\exp\{-i\kstar a\}}{z}}, \quad  
U^{+}_2 \sim \frac{C_3}{\sqrt{z}}, 
\]
as $z \to \infty$ in the lower and upper half-planes correspondingly, and $C_1,C_2,C_3$ are some unknown constants.

Next, let us consider a diffraction problem with a point source in the vicinity of the left edge of the strip (see~Figure~\ref{fig:edge_Green}, left). This corresponds to the edge Green's function satisfying the following inhomogeneous Helmholtz equation:
\begin{eqnarray}
\label{eq: edge_green_strip}
\Delta v_1(x,y)+ k_0^2 v_1(x,y)=\sqrt{\frac{\pi}{\epsilon}}\delta(x+a+\epsilon)\delta(y)
\end{eqnarray}
where \(\delta\) is the Dirac delta function and let $v_1(x,y)$ satisfy radiation conditions and Dirichlet boundary conditions on the strip. The problem is solved for a finite $\epsilon$ and then the limit $\epsilon\to 0$ is considered. Detailed analysis \cite{Shanin2003} shows that such a limit exists. Obviously, $v_1(x,y)$ violates the Meixner edge condition near the left edge.  Denote the edge Green's function corresponding to the source near the right edge as $v_2(x,y)$. Following the same procedure as for the plane wave problem (Appendix~\ref{app: strip}) we obtain the homogeneous Wiener--Hopf equation (\ref{eq: Wh_homo})  with the same kernel ${\rmA}(z)$ and
\[
 \begin{pmatrix}
 V^\pm_{1j}\\
 V^\pm_{2j}
 \end{pmatrix},
 \quad j = 1,2,
\]
% \begin{equation}
% \label{eq:Edge_problem_strip}
% {\rm V}^-_{(j)}(z) = {\rm A}(z){\rm V}^+_{(j)}(z), 
% \quad {\rm V}^\pm_{(j)} = 
% \begin{pmatrix}
% V^\pm_{1j}\\
%  V^\pm_{2j}
% \end{pmatrix}
% \quad j = 1,2,
% \end{equation}
\[
V^-_{1j}(z)  = \exp\{iz a\}\int_{-\infty}^{-a} \exp\{iz x\}v_j(x,0) dx ,
\]
\[
V^-_{2j}(z)  = \exp\{iz a\}\int_{-a}^{a} \exp\{iz x\}\frac{\ptl v_j}{\ptl y}(x,0) dx,
\]
\[
V^+_{1j}(z) = \exp\{-iz a\}\int_a^\infty \exp\{iz x\}v_j(x,0)dx, 
\]
\[
V^+_{2j}(z)  = \exp\{-iz a\}\int_{-a}^{a} \exp\{iz x\}\frac{\ptl v_j}{\ptl y}(x,0) dx.
\]
Note, since $v_j(x,y)$ do not satisfy Meixner conditions integrals $V^\pm_{2j}$ diverge near $a$ or $-a$. The regularization is done by deforming the initial contour of integration and then considering a limit (see~Appendix~\ref{app: halfline}).  
%There is no forcing term in (\ref{eq:Edge_problem_strip}), since there are no incident waves in the original problems. 
Additionally, due to Watson's lemma and the Dirac delta function in \eqref{eq: edge_green_strip}
spectral functions $V^\pm_{ij}$ are allowed to have a stronger growth at infinity \cite{Shanin2003}:
\[
V^{-}_{j1} \sim \sqrt{\frac{{i}}{{z}}}, \quad
V^{-}_{j2} \sim \delta_{j,2}{i^{3/2}}{\sqrt{z}},
\quad
V^{+}_{j2} \sim -\sqrt{\frac{{i}}{{z}}},\quad 
V^{+}_{j1} \sim -\delta_{j,2}{i^{3/2}}{\sqrt{z}},
\]
as $z \to \infty$ in the lower and upper half-planes correspondingly.
Finally, to show that a system of ${\rm V}_{(1)}$ and ${\rm V}_{(2)}$ is normal let us study
\[
{\rm X}^\pm =({\rm V}^\pm_{(1)},{\rm V}^\pm_{(2)}).
\]
It follows from (\ref{eq: Wh_homo}) and basic properties of the determinant that $\rm det\left(X\right)$ satisfies the following functional equation:
\begin{equation}
\label{eq:det_WH}
\rm det\left(\rm X^-\right) = det\left(\rm A\right)det\left(\rm X^+\right).   
\end{equation}

Taking into account that
\[
\rm det\left(\rm A\right) = -1,
\]
and the growth conditions we have 
\[
\rm det\left(\rm X^-\right) =-\rm det\left(\rm X^+\right) = -1,
\]
i.e. it is non-zero everywhere, and thus the system ${\rmV_{(1)}},{\rmV_{(2)}}$ is normal.
\begin{rem}
There is an alternative way to formulate homogeneous Wiener--Hopf problems without the limiting procedure for the point source.   One can define $v_1(x,y)$ as a solution of the homogeneous Helmholtz equation that satisfies boundary conditions, radiation condition, and has the necessary asymptotic behaviour in the vicinity of the edges. For the strip problem, we have the following estimations near the edges: 
\[
v_i(r_i,\theta_i) = -\frac{1}{\sqrt{\pi}}r_i^{-1/2}\sin(\theta_i/2) + O(r_i^{3/2}),
\]
$r_i,\theta_i$ are local coordinates in the vicinity of $(-a,0)$ and $(a,0)$. It is worth mentioning that the existence of the solution is not guaranteed for an arbitrary asymptotic behaviour, however, the necessary asymptotic estimates are known for a wide range of elliptic problems with corner points at the boundary \cite{Nazarov1994}. 
\end{rem}

\section{General procedure}
In this section, we will construct embedding formulas for the Wiener--Hopf unknowns. 
We describe two procedures for finding the canonical embedding and the plane wave embedding.

\subsection{Canonical embedding}
\label{sec: can_embed}

 The key property that we will be using to analyse \eqref{eq: matrix_WH} is that normal matrix ${\rmX^{\pm}}$ provides a multiplicative factorisation of ${\rmA}(t)$ since
 \[
 {\rmA}(t) = {\rmX}^-(t)({\rmX}^+(t))^{-1}.
 \]
 Then, by rearranging the equation \eqref{eq: matrix_WH} as usual and applying Liouville's theorem obtain (see Appendix~\ref{app: W-H})
 \begin{equation}
 \label{eq: WH_spliting}
{\rm U}^{\pm}(z,z_i) = {\rm X}^{\pm}(z)\left({\rm P}(z)\mp [\left({\rm X}^-(z)\right)^{-1}{\rm F}(z,z_i)]_{\pm}\right) 
 \end{equation}
   where
   \[\left({\rm X}^-(z)\right)^{-1}{\rm F}(z,z_i)=[\left({\rm X}^-(z)\right)^{-1}{\rm F}(z,z_i)]_-+[\left({\rm X}^-(z)\right)^{-1}{\rm F}(z,z_i)]_+,\]
 is the additive Wiener--Hopf splitting which can be computed using a Cauchy type integral,  and ${\rm P}(z)$ is a polynomial vector-function with arbitrary coefficients, which order is determined by the growth rate of $\rmU^\pm$. Normally, ${\rm P}(z)$ do not appear in diffraction-related problems, since the solution should be unique in that case.    
 
% When the forcing is a pole 
% \begin{equation}
%   \label{pole1}
% F_i(z) = \frac{{\rm r}}{z-z_i},
% \end{equation}
The case of pole forcing (\ref{eq:pole_forcing}) leads to considerable simplifications:
\begin{equation}
\label{eq:Can_solution_pole2}
{\rmU}^{+}(z,z_i) = {\rmX}^{+}(z){\rm P}(z)-\frac{{\rm X}^{+}(z)({\rmX}^-(z_i))^{-1}{\rm r}}{z-z_i}. 
\end{equation}
This equation is written as a combination of functions not dependent on \(z_i\) and constants which depend on \(z_i\). In other words, it is an embedding formula for the Wiener--Hopf equation which does not involve any Cauchy integration. This procedure can be generalised to any forcing imposed by a rational function. 

\subsection{Plane wave embedding}

In plane wave embedding the solution of a particular plane wave is expressed in terms of a fixed set of plane wave solutions. Plane wave embedding may not seem as natural as canonical one, but it has advantages from a practical point of view. Indeed, there is no need to introduce auxiliary solutions to build the normal matrix, which as in the case of edge Green's functions can have an oversingular behaviour that can lead to computational issues. In this section we will derive a plane wave embedding by following the Wiener--Hopf procedure outlined in the  Appendix~\ref{app: W-H}. An alternative derivation based on normal solutions is provided in Appendix~\ref{app: alter_emb}. 

Consider \eqref{eq: matrix_WH} with forcing (\ref{eq:pole_forcing}) as a family of Wiener--Hopf equations parameterised by parameter \(z_i\), with \(z_i\in \mathbb{R}\).
% \begin{equation}
%   \label{W-H-family}
% {\rm U}^-(z, z_i)= {\rm K}(z){\rm U}^+(z, z_i)- \frac{{\rm r^-}(z)}{(z-z_i)}.
% \end{equation}
Suppose the solution is found for fixed \(M\) such equations \eqref{eq: matrix_WH} (for how to pick them see Remark~\ref{howto_choose}) without loss of generality denote them as \(i= 1, \dots,M\).

The procedure for finding the plane wave embedding for an arbitrary \(i=\star\) in terms of the fixed \(M\) solutions  can be summarised as follows:
\begin{enumerate}
  \item Suppose the existence of the (unknown) factorisation of the matrix kernel \(\rm K=({\rm K^-})^{-1}{\rm K^+}\) then multiplying \eqref{eq: matrix_WH} by \({\rm K^-}\) obtain
    \begin{equation*}
{\rm K^- }(z){\rm U}^-(z, z_i) + {\rm K^- }(z)\frac{{\rm r^-}(z)}{(z-z_i)} = {\rm K^+}(z){\rm U}^+(z, z_i). \quad i=\star, 1, \dots,M
\end{equation*}
\item Multiply by a polynomial \(p(z,z_i)=(z-z_i)l_i(z)\), 
     \begin{equation*}
p(z,z_i){\rm K}^-(z){\rm U}^-(z, z_i)+ l_i(z){\rm K^-}(z){\rm r^-}(z) = p(z,z_i){\rm K}^+(z){\rm U}^+(z, z_i) .
\end{equation*}
\begin{rem}
In simple case it is enough to take \(p(z,z_i)=(z-z_i)\) but in cases where an additional mapping was used to derive the Wiener--Hopf equation it can be beneficial to take \(l_i(z)\) not equal to one, this will be discussed in Section~\ref{map}.
\end{rem}
\item Applying  Liouville's theorem obtain 
  \begin{equation}
   \label{canemb1}
    p(z,z_i){\rm K}^-(z){\rm U}^-(z, z_i)+ l_i(z){\rm K^-}(z){\rm r^-}(z) ={\rm Q}_i(z)         
  \end{equation}
 \begin{equation}
   \label{canemb2}
    p(z,z_i){\rm K}^+(z){\rm U}^+(z, z_i) ={\rm Q}_i(z) \quad i=\star, 1, \dots,M
  \end{equation}
  where \({\rm Q}_i(z)\) is vector of polynomials of degree \(L\), where \(L\) can be found from the known growth at infinity see Appendix~\ref{app: W-H}.
\item Now if \({\rm Q}_i(z)\) \(i= 1, \dots,M\) form a basis for the space of polynomial vectors  up to degree \(L\) (see Remark~\ref{howto_choose}) then in particular it means that there exist \(A_i\) so that \({{\rm Q}_\star(z)}=\sum_i^MA_iQ_i(z)\) hence

 \begin{equation*}
p(z,z_*){\rm U}^+_\star(z,z_*)=\sum_i^MA_ip(z,z_i){\rm U}^+(z, z_i).
\end{equation*} 
And this is an embedding formula since the functions  \(p(z,z_i){\rm U}^+(z, z_i) \) are not dependant on \(z_\star\) but the constants \(A_i\) are \(z_\star\) dependant.
\end{enumerate}
\begin{rem}
Note that a type of canonical embedding can be obtained by rearranging (\ref{canemb1}-\ref{canemb2}) for \(i=\star\).
 \end{rem}  
\begin{rem}
  \label{howto_choose}
  It is now clear that the choice of \(M\) equations from \eqref{eq: matrix_WH}  needs to be such that \({\rm Q}_i(z)\) \(i= 1, \dots,M\) form a basis for the space of polynomial vectors (of size \(N\))  up to degree \(L\). In particular, we require that \(M=LN\)  and that they are linearly independent means there does not exist \(c_i\) so that \(\sum_i^Mc_iP(z,z_i)=0\). These conditions can be checked for the given equations. In all the problems considered below it is sufficient to take different angles of incidence.
  \end{rem}

\section{Examples}
\label{sec: examples}
Below we consider three classical examples for which the Wiener--Hopf equations can be derived relatively easily. They are the problem of diffraction by a half-plane, the problem of diffraction by a finite strip, and the problem of diffraction by a wedge. Using the general procedure introduced we derive embedding formulas for these problems and show that they are consistent with the previous results.  

\subsection{Diffraction by a half-plane}
\begin{figure}[ht]
 	\centering{
 	\includegraphics[width=0.5\textwidth]{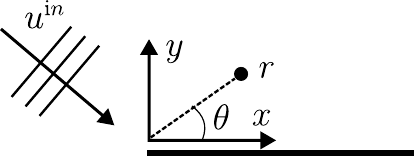}
 	}
 	\caption{Geometry of the problem of diffraction by a half-plane, with a plane wave incidence at angle $\theta_i$.}
 	\label{fig:half_plane}
 \end{figure}

% \subsubsection{Problem statement}
 Consider a classical problem of diffraction by a half-plane with Dirichlet boundary conditions, Figure~\ref{fig:half_plane}. 
Let the total field $u$ satisfy the Helmholtz equation (\ref{eq:Helm})
% \begin{equation}
% \label{eq:Helm}
% (\Delta +k^2)u(x,y) = 0
% \end{equation}
 everywhere except the half-line:
 \[
 u(x,0) = 0, \quad x\geq 0.
 \]
 The total field is presented as a sum (\ref{eq:inc_field}), and also, let $\pi>\theta_i>\pi/2$.
% Here $k$ is a wavenumber.
% Let the total field presented as a sum of incident wave $u^{\rm in}$ and scattered wave $u^{\rm sc}$:
% \begin{equation}
% \label{eq:inc_field}
% u(x,y) = u^{\rm in} + u^{\rm sc}, \quad u^{\rm in}(x,y) = \exp\{-ikx\cos\theta_i - iky\sin\theta_i\},
% \end{equation}
% and let the radiation condition be imposed in terms of limiting absorption principle, i.e. let $k$ have a small positive imaginary part. Also, let $\pi>\theta_i>\pi/2$

% Let us also introduce directivity of the scattered field:
% \begin{equation}
% \label{eq:Direct_sc}
% u^{\rm sc}(r,\theta) = \frac{\exp\{ikr-i\pi/4\}}{\sqrt{2\pi k r}}S(\theta,\theta_i)+O\left(\frac{1}{r}\right),
% \end{equation}
% where we have introduced polar coordinates:
% \[
% x = r\cos\theta,\quad y= r\sin\theta.
% \]
Additionally, there are the radiation~\cite{tikhonov1948radiation} and the Meixner edge conditions. The latter is important for the derivation of canonical embedding so is written out explicitly:
\begin{equation}
\label{Meixner_cond}
 u^{\rm sc} = Cr^{1/2}\sin(\theta/2) + O(r^{3/2}).
\end{equation}
Standard considerations show (see Appendix~\ref{app: halfline}) that the problem of diffraction by half-line can be reduced to the following Wiener--Hopf equation:
\begin{equation}
\label{eq: Wh_half_plane}
U^-(z,z_i)   =\frac{1}{i\gamma(z)} U^+(z,z_i)+ \frac{i}{z - \kstar}, \quad \kstar = k\cos\theta_i
\end{equation}
where
\begin{equation}
\label{eq:a_coef_1}
U^-(z,z_i)  = \int_{-\infty}^{0} \exp\{iz x\}u^{\rm sc}(x,0) dx,\quad U^+(z,z_i)  = \int_{0}^{\infty} \exp\{iz x\}\frac{\ptl u^{\rm sc}}{\ptl y}(x,0) dx,
\end{equation}
and $U^-(z,z_i)$ and $U^+(z,z_i)$  are analytical in the lower and the upper half-planes correspondingly. 
The Wiener--Hopf problem is supported with the following growth conditions via Watson's lemma~\cite{Olver2014}:
\[
U^{-}(z,z_i) \sim \frac{C_1}{z^{3/2}}, \quad
U^{+}(z,z_i) \sim \frac{C_2}{\sqrt{z}}, 
\]
as $z\to\infty$ in the lower and upper half-plane correspondingly, and $C_1,C_2$ are some unknown constants. Thus $U_-$ and $U_+$ tend to zero at infinity. 

Equation \eqref{eq: Wh_half_plane} is scalar and so can be easily solved using Cauchy integral, however, we will follow the procedure introduced in Section~\ref {sec: can_embed}, to illustrate the technique on a simple example.  

Consider the homogeneous version of equation \eqref{eq: Wh_half_plane}
\begin{equation}
\label{green_func_eq}
V^-(t) = \frac{1}{i\gamma(t)}V^+(t), \quad t \in R
\end{equation}
where $V^+(z)$ is analytical in upper half plane, and $V^-(z)$ is analytical in lower half-plane. Let $V^\pm(z)$ satisfy the following conditions at infinity:
\[
V^{-}(z) \sim \sqrt{\frac{i}{z}},\quad
V^{+}(z) \sim i\sqrt{iz} 
\]
as $z\to\infty$ in the lower and the upper half-plane correspondingly.
The solution of the homogeneous problems can be easily guessed:
\begin{equation}
V^{-}(z) = \frac{i\sqrt{i}}{\sqrt{k-z}},\quad  V^{+}(z) = -\sqrt{i}\sqrt{k+z}.
\end{equation}
$V^\pm(z)$ defines a function that does not turn into zero in the whole complex plane and is a normal solution of the scalar Wiener--Hopf equation. 
Thus, using (\ref{eq:Can_solution_pole2}) we can write down the solution to the inhomogeneous problem:
\begin{equation}
\label{hp_embed}
U^+(z,z_i) = -\frac{iV^+(z)(V^-(z_i))^{-1}}{z-\kstar}, 
\end{equation}

The auxiliary problem can be linked with the edge Green's function of a half-plane.  Indeed, consider the Helmholtz equation with a point source:
\begin{equation}
\label{eq:edge_greens_plane}
\Delta v(x,y) + k^2 v(x,y) = \sqrt{\frac{\pi}{\epsilon}}\delta(x - \epsilon)\delta(y),
\end{equation}
and let $v(x,y)$ satisfy Dirichlet boundary conditions on the half-line.
Then, analogously to the plane wave problem  one can obtain a functional equation (\ref{green_func_eq}) with
\[
V^-(z)  = \int_{-\infty}^{0} \exp\{iz x\}v(x,0) dx,\quad V^+(z)  = \int_{0}^{\infty} \exp\{iz x\}\frac{\ptl v}{\ptl y}(x,0) dx,
\]
where the integrals are understood in a regularised sense (see~Appendix~\ref{app: halfline}).
Introduce the directivity $S_1(\theta)$ of the edge Green's function in the same way as in (\ref{eq:Direct_sc}). Now, expressions (\ref{hp_embed}) can be rewritten in terms of the directivities.   
% \[
% v(r,\theta) = \frac{\exp\{ikr-i\pi/4\}}{\sqrt{2\pi k r}}S_1(\theta)+O\left(\frac{1}{r}\right),
% \]
Using a well-known expression that relates the spectrum of the field on the surface with the directivity of the scattered field~\cite{Shanin2003,Shanin2015}: 
\begin{equation}
S(\theta,\theta_i) = iU^+(-k\cos\theta,k\cos\theta_i), \quad S_1(\theta) = iV^+(-k\cos\theta).
\end{equation}
and the symmetry relation $1/V^-(z) = V^+(-z)$,
we obtain
\begin{equation}
\label{embed_str}
S(\theta_i,\theta) = \frac{S_1(\theta)S_1(\theta_i)}{k\cos\theta+k\cos\theta_i}.
\end{equation}
which is the well-known canonical embedding formula for the half-plane problem \cite{Noble1958,Craster2003}.
% Equation (\ref{embed_str}) is known as strong embedding formula for directivities, and is usually obtained directly from Helmholtz equation by applying a linear operator that removes the incident wave, but makes field oversingular in the vertex \cite{}.

\subsection{Diffraction by a strip}
\label{sec:strip}

% The geometry of the problem is shown in Figure~\ref{fig:strip}, and the statement is given in the introduction. It can be reduced to the $2\times 2$ matrix Wiener--Hopf problem (\ref{eq:strip_WH}), see Appendix~\ref{app: halfline},  and a system of normal solutions can be obtained from the problems for the edge Green's functions. Thus, following the procedure of section~\ref{sec: can_embed} 
The geometry of the problem is shown in Figure~\ref{fig:strip}, and the statement is given in the introduction. Using the Wiener--Hopf formulation for plane wave and edge Green's functions problems, and the embedding procedure described in Section~\ref{sec: can_embed}
we immediately obtain the canonical embedding formula:
\begin{equation}
\label{eq:strip_can_sol_+}
{\rmU}^+(z,z_i) = -\frac{{\rmX}^+(z)({\rmX}^-(\kstar))^{-1}{\rm r}}{z - \kstar}, \quad {\rm r} = i\exp\{i\kstar a\}\begin{pmatrix}
1\\0
\end{pmatrix},
\end{equation}

Let us demonstrate that from (\ref{eq:strip_can_sol_+}) follows the strong embedding formula for directivities (\ref{eq:Embed_strip_dir}). 

Again, the spectral functions are linked with the Derictivities  by the following relations:
\[
S(\theta,\theta_i) =  ie^{\mp ika}U_2^\pm(-k\cos\theta,k\cos\theta_i),
\]
\[
S_1(\theta) = ie^{\mp ika}V_{21}^\pm(-k\cos\theta),\quad S_2(\theta) = ie^{\mp ika}V_{22}^\pm(-k\cos\theta), 
\]
where  $S_1(\theta)$, $S_2(\theta)$  are directivities of the edge Green's functions with sources in vertices $(-a,0)$ and $(a,0)$, correspondingly.  Also, due to the symmetry of the problem
\begin{equation}
\label{eq:symm_strip}
V_{21}^\pm(-z) = V_{22}^\mp(z).
\end{equation}
Thus, studying the second row of (\ref{eq:strip_can_sol_+}), using the expressions for directivities, and the symmetry (\ref{eq:symm_strip})  we obtain (\ref{eq:Embed_strip_dir}).
% \begin{equation}
% \label{eq:Embed_strip_dir}
% S = \frac{S_2(\theta_i)S_2(\theta) - S_1(\theta_i)S_1(\theta)}{k\cos\theta + k\cos\theta_i},
% \end{equation}
% where we took into account {\color{red}(one need to prove it)} that
% \begin{equation}
% {\rm det}({\rm X}^-(\kstar)) = 1.
% \end{equation}

\subsection{Diffraction by a right-angled wedge}
\label{sec: wedge}
\begin{figure}[ht]
	\centering{
	\includegraphics[width=0.4\textwidth]{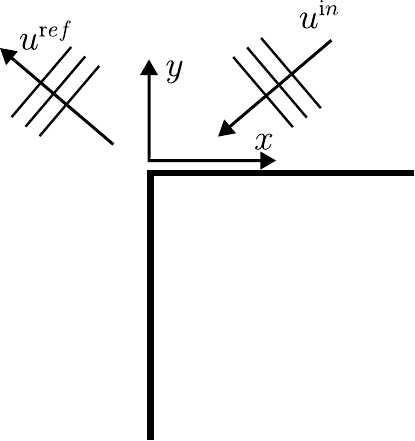}
	}
	\caption{Geometry of the problem of diffraction by a right-angled wedge.}
	\label{fig:wedge_right}
\end{figure}

Consider the problem of diffraction by a right-angled wedge, Figure~\ref{fig:wedge_right}. 
For the sake of convenience of the Wiener--Hopf formulation the total field is split in the following way:
\begin{equation}
u = u^{\rm sc}(x,y) + u^{\rm in}(x,y) - u^{\rm ref}(x,y), \quad u^{\rm ref}(x,y) = u^{\rm in}(x,-y),
\end{equation}
and restrict the angle of incidence $\pi/2>\theta_i>0$. The case $\pi>\theta_i>\pi/2$ can be studied similarly.
Then, let the incident and reflected field exist only in the upper half-plane $y>0$. To have the continuity of the total field we need to extract the geometrical optics component of the field \cite{Noble1958}. Namely, let the scattered field satisfy (\ref{eq:Helm}) and the 
following boundary conditions:
\[
u^{\rm_{sc}}(x,0) =0, \quad x>0, \quad u^{\rm_{sc}}(y,0) =0, \quad y>0,
\]
\[
u^{\rm_{sc}}(x,+0)- u^{\rm_{sc}}(x,-0)=0,
\]
\[
\frac{\ptl u^{\rm_{sc}}}{\ptl y}(x,+0)- \frac{\ptl u^{\rm_{sc}}}{\ptl y}(x,-0)=2ik\sin\theta_i\exp\{-ikx\cos\theta_i\} , \quad x<0.
\]
Such a choice of the incident and reflected fields gives us a simple expression for the directivity of the scattered filled as we are going to show below. The statement should be supported with radiation and Meixner tip conditions:
\[
u^{\rm sc}=Cr^{2/3}\sin(2/3\theta)+O(r^{4/3}).
\]

There are two Wiener--Hopf related approaches available for the problem. One involves the formulation of a pair of functional equations for the problem, which are then reduced to a scalar Wiener--Hopf equation by a carefully chosen transform of variables \cite{shanin1998excitation}. We discuss it below in Section~\ref{map}. Here, however, we are going to follow another route and formulate a matrix Wiener--Hopf problem. It is uncommon to formulate such wedge problems as matrix equations, the authors became aware of one recent work \cite{Aitken2024} after the current work was almost over. A novel technique is introduced and described in detail for a right angle wedge but which can also be applied to the wedge of an arbitrary rational angle, see Remark~\ref{rem: arb_angle}. The idea of the method is to use the reflection principle to reduce diffraction problems on the plane with straight Dirichlet scatterers to a diffraction problem on a manifold with no scatterers. This idea was famously used by Sommerfeld to solve the half-plane problem~\cite{Sommerfeld1896} and since then extended for more complicated geometries~\cite{Babic2008,malyuzhinets1955radiation,Lyalinov2005,shanin2020sommerfeld,Shanin2022}. Here the reflection principle is applied to only one face of the wedge resulting in a scattering problem on a manifold with a boundary. 

\begin{figure}[ht]
	\centering{
	\includegraphics[width=1\textwidth]{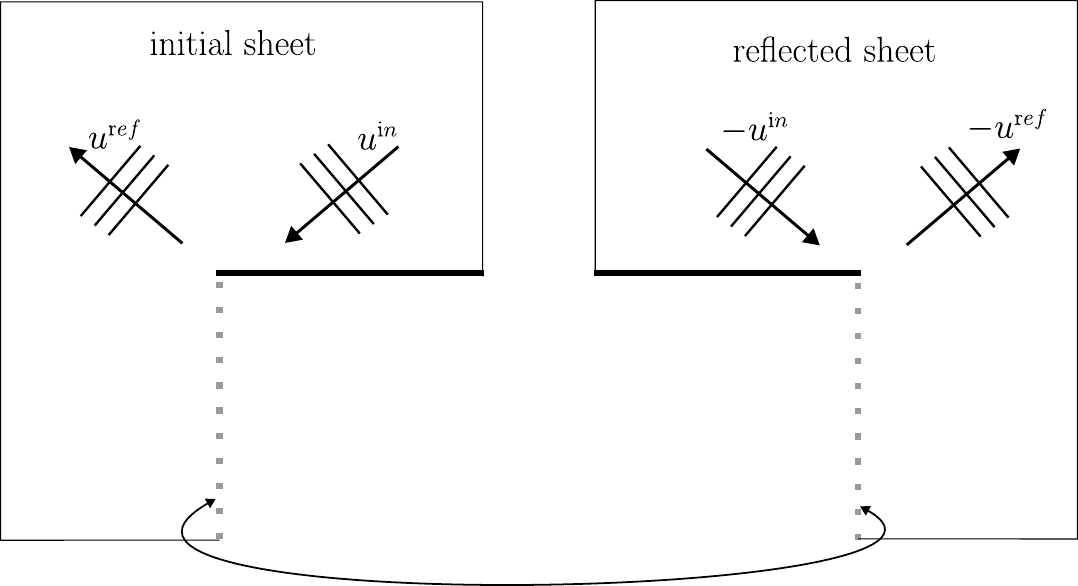}
	}
	\caption{Geometry of the problem in the reflected domain}
	\label{fig:wedge_reflected}
\end{figure}

\begin{figure}[ht]
	\centering{
	\includegraphics[width=1\textwidth]{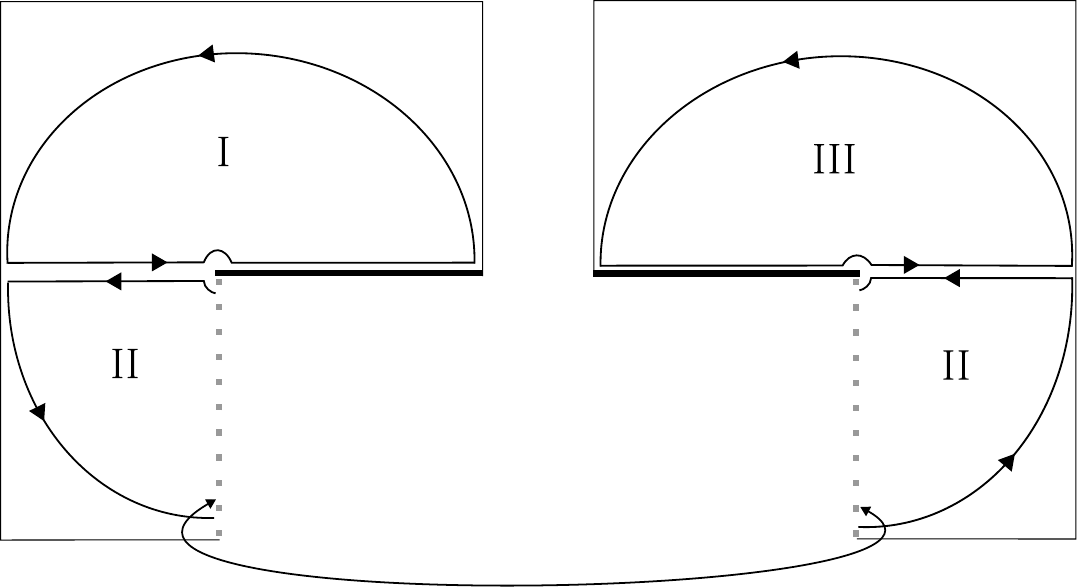}
	}
	\caption{Integration contours for Green's second identity}
	\label{fig: wedge_Greens}
\end{figure}

Let us apply the reflection principle to the face $y<0,x=0$. The result is shown in Figure~\ref{fig:wedge_reflected}.
Note, that  in the reflected domain, the boundaries lie only on $x$-axis.  Let us apply second Green's identity (\ref{eq: Greens_theorem}) to $u^{\rm sc}$ and $w_{1,2}$ with contours of integration $\Omega_i, \quad i =1\ldots 3$ covering  upper and lower half-planes (see~Figure~\ref{fig: wedge_Greens}), where 
\[
w_1 = \exp\{iz x + i\gamma(z) y\},\quad
w_2 = \exp\{iz x - i\gamma(z) y\},
\]
in the upper and the lower half-planes correspondingly.
Denote the scattered field in the vicinity of $x$-axis on the ``initial'' sheet (left in Figure~\ref{fig:wedge_reflected})  as $u^{\rm sc}_1$, and on the ``reflected" sheet as $u^{\rm sc}_2$. 
We arrive to the following matrix relation:
\begin{equation}
\renewcommand{\arraystretch}{2}
\label{wedge_WH_matrix_1}
\begin{pmatrix}
-1&0&0 \\
0 & 1 & i\gamma \\
0 & -1 & i\gamma
\end{pmatrix}
\begin{pmatrix}
W^-_2\\
W^-_1\\
\tilde U^-_1
\end{pmatrix}+
\begin{pmatrix}
0&-1&i\gamma \\
0 & 1 & i\gamma \\
-1 & 0 & 0
\end{pmatrix}
\begin{pmatrix}
W^+_1\\
W^+_2\\
\tilde U^+_2
\end{pmatrix} = \begin{pmatrix}
\dfrac{2k\sin\theta_i}{z+ \kstar}\\
0\\
\dfrac{2k\sin\theta_i}{z- \kstar}
\end{pmatrix},
\end{equation}
where 
\[
\tilde U^-_i (z,z_i)= \int^{0}_{-\infty} \exp\{iz x\} u^{\rm sc}_i(0,x) dx,\quad W^-_i (z,z_i)= \int^{0}_{-\infty}\exp\{iz x\} \frac{\partial u^{\rm sc}_i}{\partial y}(-0,x) dx,
\]
\[
\tilde U^+_i (z,z_i)= \int_{0}^{\infty} \exp\{iz x\}u^{\rm sc}_i(0,x) dx,\quad W^+_i (z,z_i)= \int_{0}^{\infty}\exp\{iz x\} \frac{\partial u^{\rm sc}_i}{\partial y}(-0,x) dx.
\]
The latter is the matrix Wiener--Hopf equation that can be re-written in the form (\ref{eq: matrix_WH}) with
% \begin{equation}
% \label{WH_matrix_wedge}
% {\rm U}^- + {\rm A}{\rm U}^+ = {\rm F},  
% \end{equation}
\[
{\rmU}^- (z,z_i)= 
\begin{pmatrix}
W^-_2(z,z_i)\\
W^-_1(z,z_i)\\
\tilde U^-_1(z,z_i)
\end{pmatrix},
\quad
{\rmU}^+ (z,z_i)=
\begin{pmatrix}
W^+_1(z,z_i)\\
W^+_2(z,z_i)\\
\tilde U^+_2(z,z_i)
\end{pmatrix}
\]   
\[
{\rmA} = \frac{i}{2\gamma}
\begin{pmatrix}
0 &2i\gamma&2\gamma^2\\
i\gamma & i\gamma& -\gamma^2 \\
-1 & 1 & i\gamma 
\end{pmatrix},\quad
\renewcommand{\arraystretch}{2}
\rmF = -
\begin{pmatrix}
\dfrac{2k\sin\theta_i}{z+ \kstar}\\
\dfrac{k\sin\theta_i}{(z- \kstar)}\\
\dfrac{ik\sin\theta_i}{(z- \kstar)\gamma}
\end{pmatrix}.
\]
Note, that forcing is not a meromorphic function, however, we will show below that due to its special structure, we are still able to derive an embedding formula without any Cauchy integrals involved.

Then, using  ideas of \cite{Aitken2024} let us multiply the equation by the following projection matrix:
\begin{equation}
{\rm P} = 
\begin{pmatrix}
1&2&0\\
-1&1&0\\
0&0&1.
\end{pmatrix},
\end{equation}
and introducing new variables
\[
{\rmU^-} \to {\rm P  U^-}=
\begin{pmatrix}
2W_1^- + W_2^-\\
W_1^- - W_2^-\\
\tilde U_1^-
\end{pmatrix},
{\rmU^+} \to {\rm P  U^+}=
\begin{pmatrix}
2W_2^+ + W_1^+\\
W_2^+ - W_1^+\\
\tilde U_2^+
\end{pmatrix},
\]
obtain Wiener--Hopf equation (\ref{eq: matrix_WH}) with 
\begin{equation}
{\rmA} \to {\rm P  \rmA P^{-1}}=\frac{1}{2\gamma} 
\begin{pmatrix}
-1&0&0\\
0&\gamma&-3i\gamma^2\\
0&i&-\gamma.
\end{pmatrix},
\quad
{\rmF} = \frac{-k\sin\theta_i}{(z-\kstar)\gamma}\begin{pmatrix}
 2\gamma\\
 \gamma\\
i
\end{pmatrix} +
\frac{2k\sin\theta_i}{(z+\kstar)} 
\begin{pmatrix}
-1\\
1\\ 
0
\end{pmatrix}.
\end{equation}
Hence, the problem is reduced to a scalar additive factorization problem, and $2\times 2$ matrix Wiener--Hopf problem. Let us concentrate on the latter (one can easily show that the solution of this sub-problem gives the solution of the original diffraction problem). Finally, we have $2\times 2$ equation $(\ref{eq: matrix_WH})$ with
\[
{\rmU^-}=
\begin{pmatrix}
W_1^-- W_2^-\\
\tilde U_1^-
\end{pmatrix},
{\rmU^+}=
\begin{pmatrix}
W_2^+ - W_1^+\\
\tilde U_2^+
\end{pmatrix},
\]
\begin{equation}
\label{eq: WH_edge_final}
{\rmA}=
\frac{1}{2\gamma}
\begin{pmatrix}
\gamma&-3i\gamma^2\\
i&-\gamma
\end{pmatrix},
\quad
{\rmF} = \dfrac{-k\sin\theta_i}{(z-\kstar)\gamma}\begin{pmatrix}
 \gamma\\
i
\end{pmatrix} +
\dfrac{2k\sin\theta_i}{(z+\kstar)} 
\begin{pmatrix}
1\\
0
\end{pmatrix}.
\end{equation}
Unknown functions satisfy the following growth conditions: 
\[
\tilde U^{-}_1\sim \frac{C_1}{z^{5/3}}, \quad
W^{-}_i  \sim \frac{C_2}{z^{5/3}}, \quad
\tilde U^{+}_2 \sim  \frac{C_3}{z^{5/3}}, \quad
W^{+}_i\sim  \frac{C_4}{z^{5/3}}, 
\]
as $z\to\infty$ the lower and upper half-planes correspondingly, and $C_1,\ldots,C_4$ are some unknown constants.

\begin{figure}[ht]
	\centering{
	\includegraphics[width=0.8\textwidth]{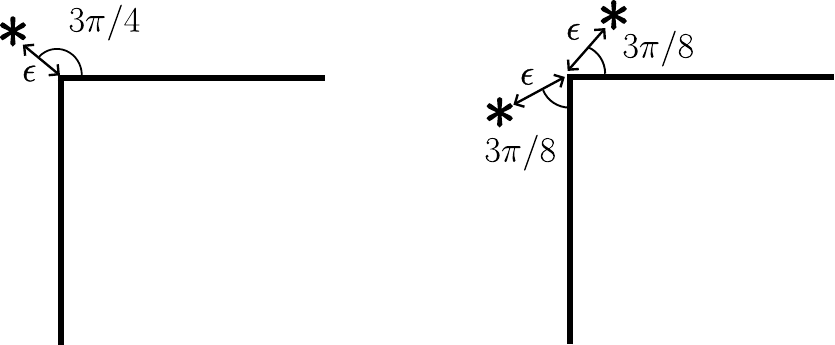}
	}
	\caption{Geometry of problems for edge Green's functions}
	\label{fig:wegde_greens_geom}
\end{figure}

 Let us introduce auxiliary problems for edge Green's functions (see Figure~\ref{fig:wegde_greens_geom}).
 Namely, let $v_1$ satisfy the following inhomogeneous Helmholtz equation: 
\[
(\Delta v_1 + k^2 v_1) = -\frac{\pi}{\epsilon^{2/3} r} \delta(r-\epsilon)\delta(\theta - 3\pi/4),
\]
and $v_2$ satisfy 
\[
(\Delta v_2 + k^2 v_2) = \frac{\pi}{\epsilon^{4/3}r} \left[\delta(r-\epsilon)\delta(\theta - 9\pi/8)-\delta(r-\epsilon)\delta(\theta - 3\pi/8)\right].
\]
Both functions should also satisfy the radiation conditions and the Dirichlet boundary conditions on the faces of the wedge. Such a choice of source combination is motivated by the structure of the Meixner series in the vicinity of the edge, and we refer the reader to \cite{Craster2005} for the details. 

As was mentioned above, instead of the edge Green's functions one may consider the oversingular solutions with the following growth in the origin:
\[
v_1 \sim r^{-2/3}\sin(2/3\theta), \quad v_2 \sim r^{-4/3}\sin(4/3\theta).
\]
Applying the technique developed above we obtain a pair of homogeneous Wiener--Hopf problems (\ref{eq: Wh_homo}) for unknown functions
\[
 \begin{pmatrix}
 V^\pm_{1j}(z)\\
 V^\pm_{2j}(z)
 \end{pmatrix},
 \quad j = 1,2,
\]
\[
V^-_{1j}(z) = \int^{0}_{-\infty}\exp\{iz x\} \left(\frac{\partial v_{j1}}{\partial y}(0,x) -  \frac{\partial v_{j2}}{\partial y}(0,x)\right)dx,
\]
\[
V^-_{2j}(z) = \int^{0}_{-\infty} \exp\{iz x\} v_{j1}(0,x) dx,
\]
\[
V^+_{1j}(z) = \int_{0}^{\infty}\exp\{iz x\} \left(\frac{\partial v_{j2}}{\partial y}(0,x) -  \frac{\partial v_{j1}}{\partial y}(0,x)\right)dx,
\]
\[
V^+_{2j}(z) = \int_{0}^{\infty} \exp\{iz x\} v_{j2}(0,x) dx,
\]
and  the kernel $\rmA(z)$ given by (\ref{eq: WH_edge_final}). Here, $v_{ji}(x,y)$, is $j$-th edge Green's function taken on  $i$-th sheet.
Unknown functions satisfy the following growth conditions:
\begin{equation}
{\rm V}^\pm_{(1)}  \sim 
\begin{pmatrix}
C^\pm_{11}z^{2/3}\\
C^\pm_{21}z^{-1/3}
\end{pmatrix},\quad 
{\rm V}^-_{(2)}  \sim 
\begin{pmatrix}
C^\pm_{12}z^{4/3}\\
C^\pm_{22}z^{1/3}
\end{pmatrix},
\label{growth_cond_can_wedge} 
\end{equation}
where
\[
C^-_{11} = -\Gamma\left(-\frac{2}{3}\right),\quad
C^-_{12} = 2\Gamma\left(-\frac{4}{3}\right),
\]
\[
C^-_{21} = i\frac{\sqrt{3}}{2}\Gamma\left(\frac{1}{3}\right),\quad
C^-_{22} = i\frac{\sqrt{3}}{2}\Gamma\left(-\frac{1}{3}\right),\quad
\]
where $\Gamma(x)$ is the gamma function, and $C^+_{ij}$ can be determined from the symmetry of the problem. Indeed, from the symmetry of the edge Green's functions
\[
v_{j1}(y,x) = - v_{j2}(y,-x),
\]
% \[
% \left(\rmA(z)\right)^2 ={\rm I},\quad  \rmA(-z) = \rmA(z),
% \]
%where $\rm I$ is the identity matrix,
follows the symmetry relation
\begin{equation}
V^-_{ij}(z) = -V^+_{ij}(-z).
\label{eq: wedge_symmmetry}   
\end{equation}
Let us study the matrix 
$\hat \rmX = \left(\rmV_{(1)},\rmV_{(2)}\right).$
Using
\[
{\rm det}(\rmA) = -1, 
\]
(\ref{eq: wedge_symmmetry}) and (\ref{eq:det_WH}), we get that
\[
{\rm det} (\hat \rmX^-) = -{\rm det} (\hat \rmX^+)  = -9\pi i z,
\]
and thus $\hat \rmX$ is not normal. However, with a minor modification, it can be used in the same way as the normal matrix. Indeed, let us move the contour of Wiener--Hopf problem slightly in the lower half-plane thus excluding the origin from the domain of analyticity of ``minus'' functions. Then consider the following factorization of the kernel:
\[
 {\rmA}(z) = \left(\frac{{\hat \rmX}^-(z)}{z}\right)\left(z({\hat \rmX}^+(z))^{-1}\right).
\]
Matrix ${\hat \rmX}^-(z)/z$ is analytic in the lower half-plane with nonzero determinant everywhere, and $z({\hat \rmX}^+(t))^{-1}$ is analytic in the upper half-plane. Thus, Liouville's theorem theorem can be applied after $({\rm X}^-)^{-1}{\rm \hat F}_2$ is additively factorized.
%\begin{equation}
%({\rm X}^-_2)^{-1}{\rm \hat U}^-_2 = ({\rm X}^+_2)^{-1}{\rm \hat U}^+_2 + ({\rm X}^-_2)^{-1}{\rm \hat F}_2.
%\label{Half_factor_wedge}
%\end{equation}
Note that
\begin{equation}
\label{additive_F}
({\hat \rmX}^-)^{-1}{\rmF} = \frac{2k\sin{ \theta_i}}{z+\kstar}{(\hat \rmX^-)}^{-1}\begin{pmatrix}
1\\
0
\end{pmatrix}
-\frac{2k\sin{ \theta_i}}{z-\kstar}{(\hat \rmX^+)}^{-1}\begin{pmatrix}
1\\
0
\end{pmatrix},
\end{equation}
where we used that 
\[
{\rmA}^{-1}\begin{pmatrix}
\gamma\\
i
\end{pmatrix}=
\begin{pmatrix}
2\gamma\\
0
\end{pmatrix}.
\]
Using the latter (\ref{additive_F}) can be factorized by pole removal:
\[
({\hat \rmX}^-)^{-1}{\rmF} = \left[({\hat \rmX}^-)^{-1}{\rmF}\right]^- - \left[({\hat\rmX}^-)^{-1}{\rmF}\right]^+,
\]
where $\left[({\rm X}^-)^{-1}{\rm \hat F}\right]^-$ and $\left[({\rm X}^-)^{-1}{\rm \hat F}\right]^+$ analytic in the lower and the upper half-planes correspondingly. Then, for term analytic in the lower half-plane, we have
\[
\left[({\hat\rmX}^-)^{-1}{\rmF}\right]^-(z,z_i) =
\]
\[
\frac{2k\sin{ \theta_i}}{z+\kstar}\left[{(\hat\rmX^-(z))}^{-1} - {(\hat\rmX^-(-\kstar))}^{-1}\right]\begin{pmatrix}
1\\
0
\end{pmatrix}
-\frac{2k\sin{ \theta_i}}{z-\kstar}({(\hat\rmX^+)(\kstar)})^{-1}\begin{pmatrix}
1\\
0
\end{pmatrix}.
\]
Using the latter, after a bit of algebra, we obtain the embedding formula:
\begin{equation}
\label{Wedge_sol}
{\rm U}^-(z,z_i) = \frac{4z_i k\sin{\theta_i}}{z^2-\kstar^2}{\hat\rmX}^-(z){({\hat\rmX}^-(-\kstar))}^{-1}\begin{pmatrix}
1\\
0
\end{pmatrix}
+\frac{2k\sin{ \theta_i}}{z+\kstar}
\begin{pmatrix}
1\\
0
\end{pmatrix},
\end{equation}
where we took into account the symmetry relation (\ref{eq: wedge_symmmetry}).

% Using the same symmetry argument one can obtain
% \[
% {\rm U}^+ (z,z_i) = -{\rm U}^-(-z,z_i).
% \] 

Let us re-write (\ref{Wedge_sol}) in terms of directivities. As usual, spectral functions are linked with directivities:
\begin{equation}
S(\theta,\theta_i) = -k\sin\theta \tilde U^-_1(-k\cos\theta), \quad S_j(\theta) = -k\sin\theta V^-_{2j}(-k\cos\theta).
\end{equation}
Substituting the latter in (\ref{Wedge_sol}) we get:
\[
S(\theta,\theta_i) = \frac{4}{9\pi ik^2}\frac{S_1(\theta)S_2(\theta_i)-S_1(\theta_i)S_2(\theta)}{\cos^2\theta - \cos^2\theta_i},
\]
which is consistent with \cite{Craster2005}. Additionally, for the sake of completeness, the kernel $\rmA(z)$ is factorized in the Appendix~\ref{app:wedge} and explicit expressions for $\hat\rmX(z)$ and $S(\theta,\theta_i)$ are given.

\begin{rem}
\label{rem: arb_angle}
Let us consider a wedge with  a rational  angle
\begin{eqnarray}
\label{eq: rat_angle}
\varphi = \frac{q\pi}{p},\quad  q>p.  
\end{eqnarray}
Applying the reflection principle to one of wedge faces $p$-times in a row, we will arrive to a problem on a multi-sheeted surface with boundaries laying only on the $x$-axis. Applying Green's theorem in the same manner as it was done previously we arrive to a matrix Wiener--Hopf problem of dimension $q\times q$. Direct computations show that the matrix kernel has $q-p$ constant eigenvectors, which means that by using some constant projection matrix  $q\times q$ Wiener--Hopf problem can be reduced to  $p\times p$ Wiener--Hopf problem and $q-p$ scalar additive factorization problems. The reduced problem can then be solved using the method of the normal matrix which will lead to the canonical embedding formula for the wedge problem with a rational angle. 
\end{rem}

\section{Wiener--Hopf equations derived via additional mappings}
\label{map}

In the examples considered above, the spectral variable (variable  of the Fourier transform along $x$-axis) and the variable of Wiener--Hopf problem  coincide, this results in the embedding formulae having the following structure: 
\begin{equation}
U^\pm(z,z_i) = \frac{\sum_{n=1}^N C_n(z_i)V_n^\pm(z)}{L(z,z_i)},
\label{eq: gen_embedding}
\end{equation}
where $U^\pm(z,z_i)$ a spectral function for the problem with the plane wave,  $V_i^\pm(z)$ are spectral functions associated with the auxiliary problems, $C_i(z_i)$ known functions that are expressed in terms of $V^\pm_i(z_i)$, and $L(z,z_i)$ is a polynomial function of $z$ and $z_i$.  
These expressions were derived in~\cite{Craster2003,Biggs2006} by applying a local operator to the physical field in particular a differential operator $H$ to express the result as a finite sum of other solutions: 
\[
H[u](x,y) = \sum_{n=1}^N C_n(z_i)v_n(x,y).
\]
For example, the operator used for the finite strip and half-plane is
\begin{equation}
\label{eq: emb_op_strip}
H[u] = \frac{\ptl u}{\ptl x} + iku \cos\theta_i ,  
\end{equation}
and for the right-angled wedge is
\begin{equation}
\label{eq: emb_op_wedge}
H[u] = \frac{\ptl^2 u}{\ptl x^2} - (iku \cos\theta_i)^2.
\end{equation}

\begin{figure}[ht]
	\centering{
	\includegraphics[width=1\textwidth]{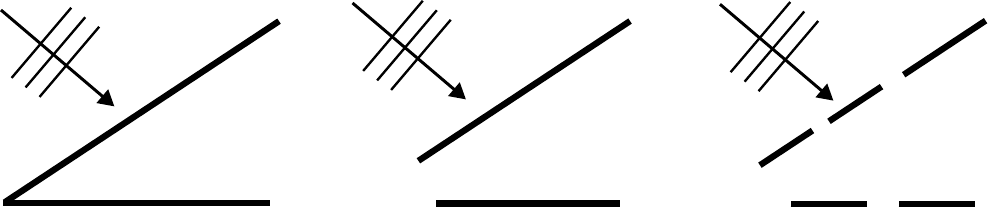}
	}
	\caption{Diffraction problems that can be reduced to Wiener--Hopf problems via additional mapping.}
	\label{fig: wedges}
\end{figure}

Let us consider a more general situation where the spectral variable can be connected to the Wiener--Hopf variable by some additional mapping: 
\begin{equation}
\label{eq: mapping}
\alpha = \eta(z).
\end{equation}
A well-known example where an additional mapping is used are problems of diffraction by a wedge-like geometry (several examples are presented in Figure~\ref{fig: wedges}). Some more sophisticated problems were considered by \cite{Daniele2020, Daniele_wedgebook20}. 
Generally, normal solutions (\ref{eq:Can_solution_pole2}) will not lead to a local relation between physical fields and lead to a more general embedding formula~\cite{Shanin2010}.
However, with some modifications to the procedure it is possible to recover the usual embedding formula in terms of other solutions. 

Suppose that there are exist a polynomial function $Q(z,z_i)$ such that
\begin{equation}
\label{eq: alg_property}
\frac{Q(z,z_i)}{z-z_i} = M(z,z_i),    
\end{equation}
where $Q(z,z_i)$ is a polynomial both in variable $z$ and variable $\alpha$ after mapping (\ref{eq: mapping}), and $M(z,z_i)$ is a polynomial of variable $z$. Using (\ref{eq: alg_property})  rewrite (\ref{eq:Can_solution_pole2}) as
% \begin{equation}
% \label{eq:Can_solution_pole3}
% {\rmU}^{-}(z,z_i) = \frac{{\rmX}^{-}(z){\rm P}(z)}{H(z,z_i)}+\frac{M(z,z_i){\rm r} -{\rmX}^{-}(z)({\rmX}^-(z_i))^{-1}{\rm r}}{H(z,z_i)}, 
% \end{equation}
\begin{equation}
\label{eq:Can_solution_pole4}
{\rmU}^{+}(z,z_i) = {\rmX}^{+}(z){\rm P}(z)-\frac{M(z,z_i){\rm X}^{+}(z)({\rmX}^-(z_i))^{-1}{\rm r}}{Q(z,z_i)}. 
\end{equation}
 Study the combination
\[
M(z,z_i){\rm X}^+(z)({\rmX}^-(z_i))^{-1}{\rm r},
\]
and without the loss of generality let $M(z,z_i)$ be a polynomial of order $L$:
\[
M(z,z_i) = \sum_{m=0}^LC_l(z_i)z^l.
\]
Then, note that if $\rmX^\pm$ is a matrix of solutions of (\ref{eq: Wh_homo}), then $z^l\rmX^\pm$ is also a matrix of solutions of (\ref{eq: Wh_homo}). Thus, the combination above can be re-written as
\[
\rmE\hat \rmX^+(z)  \rmC(z_i) (\hat\rmX^{-}(z_i))^{-1}\rmE^T{\rm r},
\]
where 
\[
\hat \rmX^{\pm} = 
\begin{pmatrix}
\rm X^\pm &\rm 0&\ldots&{\rm 0}\\
\rm 0 &z\rm X^\pm&\ldots&{\rm 0}\\
\vdots &\vdots&\ddots&\vdots\\
\rm 0 &\rm 0&\cdots&z^L\rm X^\pm\\
\end{pmatrix},\quad
 \rmC = 
\begin{pmatrix}
C_0\rm I &\rm 0&\ldots&{\rm 0}\\
\rm 0 &C_1(z_i)\rm I&\ldots&{\rm 0}\\
\vdots &\vdots&\ddots&\vdots\\
\rm 0 &\rm 0&\cdots&C_L(z_i)\rm I\\
\end{pmatrix},
\]
\[
\rmE = (\rm I,\ldots,\rm I),
\]
and $\rm I$ is the identity matrix of size $N$.
Note, that here $\hat \rmX^{\pm}(z)$ is a matrix of size $L\times N$ composed of solutions of the homogeneous Wiener--Hopf problem. Then, substituting the latter into (\ref{eq:Can_solution_pole4}) we obtain an embedding formula of the form (\ref{eq: gen_embedding}):
% \begin{equation}
% \label{eq:Can_solution_pole5}
% {\rmU}^{-}(z,z_i) = {\rmX}^{-}(z){\rm P}(z)+\frac{\rmE\hat \rmX^{-}(z)  \rmC(z_i) \hat\rmX^{-}(z_i)^{-1}\rmE^T{\rm r}}{Q(z,z_i)}, 
% \end{equation}
\begin{equation}
\label{eq:Can_solution_pole6}
{\rmU}^{+}(z,z_i) ={\rmX}^{+}(z){\rm P}(z)-\frac{\rmE\hat \rmX^{+}(z)  \rmC(z_i) (\hat\rmX^{-}(z_i))^{-1}\rmE^T{\rm r}}{Q(z,z_i)}. 
\end{equation}
Therefore, the modified embedding formulae lead to a local relation between the physical fields. 
As a trade off it requires $L\times N$ solutions of the homogeneous Wiener--Hopf problem  (unlike \eqref{eq:Can_solution_pole2} which needs $N$) but these $L\times N$ are easily constructed from the  $N$ solutions as indicated above. 

Determining the class of mappings that satisfy (\ref{eq: alg_property}) is beyond the scope of this paper. However,  such polynomials $Q(z,z_i)$ and $M(z,z_i)$ do exist for diffraction problems in angular geometries. In particular, for angular geometries having a rational angle (\ref{eq: rat_angle}), the mapping \eqref{eq: mapping} can be taken as \cite{shanin1998excitation}
\begin{equation}
\label{eq: var_change}
\alpha = \cos^2\left(\frac{p}{2q}\arccos\left(\frac{z}{k}\right)\right).
\end{equation}
% The equation in the \(z\)-variable has the form
% \begin{equation}
% \label{spectral_angular}
% {\rm U_1}(\cos(z)) = {\rm K }(z){\rm U_2}(\cos(\frac{p}{2q}\pi-z)) + {\rm F}(z),
% \end{equation}
% and the Wiener--Hopf equation in the \(z\)-plane has the form
%  \begin{equation*}
%    {\rm U_1}(\cos(z(z))) = {\rm K }{\rm U_2}(z(z)){\rm }(\cos(z(-z))) + {\rm F}(z(z)),
%  \end{equation*}  
%  where \({\rm U_1}(\cos(z(z)))/{\rm U_2}(\cos(z(-z)))\) analytic in upper/lower  \(z\)-half-plane.
Then, consider 
\begin{equation}
\label{eq:pol_Q}
Q(z, z_i)=\cos (p\theta(z))-\cos(p\theta_i(z)), \quad z = k\cos(\theta), \quad z_i = k\cos(\theta_i),     
\end{equation}
where \(\theta_i\) is the plane wave incidence angle. This is a polynomial in both $z$ and $\alpha$ which ensures the locality of the resulting embedding formula. Indeed,
\begin{equation*}
 Q(z,z_i)= \cos \left(p \arccos\left(\frac{z}{k}\right)\right)-\cos(2p\theta_i)=T_p\left(\frac{z}{k}\right)-\cos(2p\theta_i).
\end{equation*}
where \(T_p\) are the Chebyshev polynomials of the first kind \(T_n(\cos(\theta))=\cos(n\theta)\). Similarly,
\begin{equation*}
 Q(z,z_i)= \cos \left(q \arccos\left(2\alpha-1\right)\right)-\cos(2p\theta_i)=T_q(2\alpha-1)-\cos(2p\theta_i),
\end{equation*}
i.~e. $Q$ is the $q$-th Chebyshev polynomial in $\alpha$ variable, and $p$-th Chebyshev polynomial in $z$ variable.

The polynomial $Q$ will be used below to derive a local embedding formula for the problem of diffraction by a right-angled wedge.

\subsection{Wiener--Hopf equation for the right-angled wedge via additional mapping}
%Again, for the sake of simplicity, we consider the problem of diffraction by a right-angled wedge. The geometry of the problem is shown in figure~\ref{fig:wedge_right}. 

\begin{figure}[ht]
	\centering{
	\includegraphics[width=0.4\textwidth]{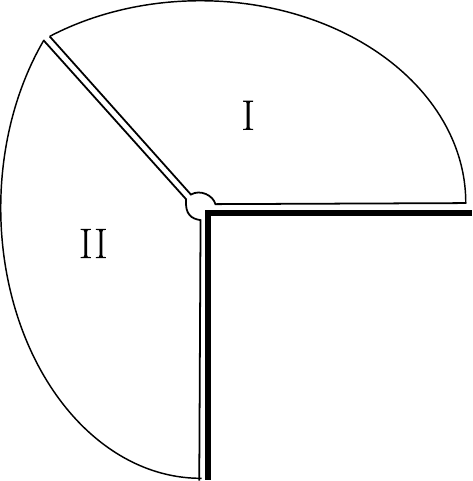}
	}
	\caption{Integration contours for Green's second identity}
	\label{fig:wedge_greens_domains}
\end{figure}

Let us slightly modify the statement given in Section~\ref{sec: wedge}. Namely, let the total field consist  only of the incident wave and the scattered field 
\[
u = u^{\rm sc}(x,y) + u^{\rm in}(x,y),
\]
which is true when the angle of incidence is in between    $\pi/2$ and $\pi$. Let us rewrite the incident wave and the outgoing wave $w$ in the polar coordinates:
\[
u^{\rm in} = \exp\{-ikr\cos(\theta - \theta_i)\},\quad w = \exp\{ikr\cos(\theta - \psi(z))\},
\]
where $z = k\cos\psi$. We cannot apply Green's identity to $u^{\rm sc}$, and $w^{\rm sc}$ in the whole domain, since the angle of the wedge is bigger than $\pi$, and there always be a set of directions along which $w$ grows exponentially. Instead, let us partition the domain into two sectors of a circle as shown in Figure~\ref{fig:wedge_greens_domains}, and apply Green's identity twice.
Obtain:
\begin{equation}
\label{eq: func_wedge_1}
S(\psi) + \Phi(\cos\psi) = \frac{-\sin\psi(z)}{\cos\psi - \cos\theta_i},
\end{equation}
\begin{equation}
\label{eq: func_wedge_2}
S(\psi) + \Psi(\cos(\psi+\varphi)) = \frac{\sin(\varphi+\psi)}{\cos(\varphi+\psi - \cos(\varphi+\theta_i)}, \quad \varphi = \frac{3\pi}{2},
\end{equation}
where 
\[
S(\psi) = \int_\Gamma \left[\frac{\ptl u^{\rm sc}}{\ptl n} w -\frac{\ptl w^{\rm sc}}{\ptl n} u \right]dl,
\]
\[
\Phi(q) = \int_{0}^{\infty}\frac{\ptl u^{\rm sc}}{\ptl n}(r,0)e^{ikrq}dr,\quad
\Psi(q) = \int_{0}^{\infty}\frac{\ptl u^{\rm sc}}{\ptl n}(r,\varphi)e^{ikrq}dr,
\]
where $\Gamma$ is a ray  $\Gamma = \rho e^{i\gamma},\quad \rho\in [0,\infty)$, $\gamma$ is a constant between $\pi/2$ and $\pi$.  Equations (\ref{eq: func_wedge_1} -\ref{eq: func_wedge_2}) can be considered as analytic continuation formulae for $S(\psi)$. Both equations are valid in the region $\psi \in [\pi/2;\pi]$. Thus, for these values of $\psi$ one can exclude $S(\psi)$:
\[
\Phi (\cos\psi) - \Psi(\cos(\psi+\varphi)) = \frac{\sin\psi}{\cos\psi - \cos\theta_i} - \frac{\sin(\varphi+\psi)}{\cos(\varphi+\psi) - \cos(\varphi+\theta_i)}.
\]
The right-hand side can be simplified with the use of the following identity:
\[
\cos\psi\sin(\varphi - \psi) + \sin\psi\cos(\varphi - \psi) =  \cos\theta_i\sin(\varphi - \theta_i) + \sin\theta_i\cos(\varphi - \theta_i).
\]
Thus, we get
\begin{equation}
\label{eq:WH_wedge_final}
\Phi(\cos\psi) - \Psi(\cos(\psi+\varphi)) = \frac{\sin\theta_i}{\cos\psi - \cos\theta_i} - \frac{\sin(\theta_i+\varphi)}{\cos(\varphi+\psi) - \cos(\varphi+\theta_i)}.
\end{equation}
Then, using let us use the change of variables (\ref{eq: var_change}) with $q=3$, $p=2$, and re-write it in terms of variable $\psi$: 
\begin{equation}
\label{eq:var_change_psi}
\psi = \frac{2q}{\pi p}\arccos(\sqrt{\alpha}).   
\end{equation}
As was shown in \cite{shanin1998excitation,Nethercote2020}, such a change of variables transforms the equation (\ref{eq:WH_wedge_final}) into the Wiener--Hopf equation.
 Indeed,
\[
\Phi (\cos\psi(\alpha)) - \frac{\sin\theta_i}{\cos\psi(\alpha) - \cos\theta_i}  = \Psi(\cos(\psi(\alpha)+\varphi))- \frac{\sin(\theta_i+\varphi)}{\cos(\varphi+\psi(\alpha)) - \cos(\varphi+\theta_i)}
\]
defines a function analytical in the whole complex plane $\alpha$, except the pole $\alpha^{\rm in} = \cos^2 \left(\theta_i/3\right)$ with the residue
\[
r(\theta_i) = \lim_{\alpha \to \alpha^{\rm in}}\left[(\alpha - \alpha^{\rm in})\frac{\sin\theta_i}{\cos\psi(\alpha) - \cos\theta_i}\right] = \frac{1}{3} \cos\left(\frac{2}{3}\left(\theta_i-\frac{3\pi}{4}\right)\right).
\]
Thus, the solution is 
\[
\Phi(\cos\psi(\alpha)) = \frac{r(\theta_i)}{\alpha - \alpha^{\rm in}}.
\]

Standard computations show \cite{Nethercote2020} that the directivity of the diffracted field can be expressed as follows: 
\begin{equation}
S(\theta,\theta_i) = -2i[\Phi(\cos(\theta-\pi)) - \Phi(\cos(\theta+\pi))]. 
\label{eq:Direct_wedge}    
\end{equation}

%\subsubsection{Embedding formula}
The derivation of embedding for equation (\ref{eq:WH_wedge_final}) reduces to multiplication by some polynomial of $\alpha$. In general, such a multiplication by polinomial will correspond to some non-local integro-differential operator applied to a field in physical space, but some specific choices will lead to a local one. For example, a natural embedding $\alpha-\alpha_0$ gives an embedding that is valid for a wedge of any angle (including irrational), but the auxiliary problem will correspond to a physical problem with a source density distributed along the wedge boundary, and the embedding operator will be given by the Green's integral. This case is discussed in detail in \cite{Shanin2010}. 

Now, let us consider the polynomial $Q$ for $p=2$. Writing it down as a function of $\alpha$ we get:
\begin{equation}
\label{eq:wedge_polynomial}
Q(z(\alpha),z_i) = 4(2\alpha-1)^3 - 3(2\alpha-1) - \cos(2\theta_i).    
\end{equation}
Note that $Q$ is linked with the Fourier transform of (\ref{eq: emb_op_wedge}).
% This polynomial corresponds to the local differential operator in the physical domain (equivalent up to a coefficient to the embedding operator defined in \cite{}). Note, that $H(\alpha,\theta_i)$ is a polynomial with respect to variables $\alpha$ and $z = \cos\psi$:
% \[
% H(\alpha(z),\theta_i) = 2z^2-1- \cos(2\theta_i),
% \]
% and it satisfies property (\ref{eq: alg_property}).
In order not to repeat the results of Section~\ref{sec: wedge}, let's derive the plane wave embedding formula for the wedge problem.
Multiply (\ref{eq:WH_wedge_final}) by 
(\ref{eq:wedge_polynomial}) and after the application of Liouville's theorem get:
\[
Q(z(\alpha),z_i)\Phi(\cos\psi(\alpha)) = c_0(\theta_i) + c_1(\theta_i)\alpha + c_2(\theta_i)\alpha^2, 
\]
where
\begin{equation}
c_0(\theta_i) = 32r(\theta_i) \cos^2\left(\frac{\theta_i}{3}-\frac{\pi}{3}\right)\cos^2\left(\frac{\theta_i}{3}-\frac{2\pi}{3}\right),   
\end{equation}
\begin{equation}
\label{eq:wedge_c_coefs1}
c_1(\theta_i) = -32r(\theta_i) \left[\cos^2\left(\frac{\theta_i}{3}-\frac{\pi}{3}\right)+\cos^2\left(\frac{\theta_i}{3}-\frac{2\pi}{3}\right)\right],   
\end{equation}
\begin{equation}
\label{eq:wedge_c_coefs2}
c_2(\theta_i) = 32r(\theta_i).
\end{equation}
Let us consider two more spectral functions  $\Phi_1,\Phi_2$ corresponding to the problems with different incident angles $\theta_1,\theta_2$ and associated spectral parameters $z_1,z_2$. Consider a linear combination
\[
R(\alpha) = Q(z(\alpha),z_i)\Phi(\cos\psi(\alpha)) - B_1Q(z(\alpha),z_1)\Phi_1(\cos\psi(\alpha))-
\]
\[
- B_2Q(z(\alpha),z_2)\Phi_2(\cos\psi(\alpha)).
\]
Since the latter is just a polynomial of the order 2, let us choose the coefficients $B_1,B_2$ in a way that cancels the terms $\alpha$ and $\alpha^2$ in this combination. Indeed,
taking $B_1,B_2$ as solutions of the following system of linear equations
\begin{equation}
\label{eq:wedge_coefsB}
\begin{pmatrix}
 c_1(\theta_1) & c_1(\theta_2)\\
 c_2(\theta_1) & c_2(\theta_2)
\end{pmatrix}
\begin{pmatrix}
 B_1\\
 B_2
\end{pmatrix}
=
\begin{pmatrix}
c_1(\theta_i)\\
c_2(\theta_i)
\end{pmatrix},    
\end{equation}
we get
% \[
% H(\alpha,\theta_i)\Phi(\cos\psi(\alpha)) - B_1H(\alpha,\theta^1)\Phi_1(\cos\psi(\alpha))- B_2H(\alpha,\theta^2)\Phi_2(\cos\psi(\alpha)) = C,
% \]
% where C is a known function of incident angles $\theta^1,\theta^2,\theta_i$, and does not depend on $\alpha$:
\[
R(\alpha) = c_0(\theta_i) - c_0(\theta_{1}) - c_0(\theta_{2}),
\]
i.e $R$ does not depend on alpha, and is only a function of incident angles $\theta_1,\theta_2,\theta_i$. Using (\ref{eq:Direct_wedge}), and (\ref{eq:var_change_psi}) we obtain embedding for  directivities:
\begin{equation}
S(\theta,\theta_i) = \frac{B_1\hat S(\theta,\theta_1) + B_2 \hat S(\theta,\theta_2)}{\cos(2\theta)-\cos(2\theta_i)},   
\end{equation}
where $\hat S$ is a weighted directivity:
\[
\hat S(\theta,\theta_j) = \left(\cos(2\theta)-\cos(2\theta_{j})\right)S(\theta,\theta_{j}).
\]
Coefficients $B_1,B_2$ is defined by relations (\ref{eq:wedge_c_coefs1}--\ref{eq:wedge_coefsB}). The result is consistent with the embedding formula for plane waves derived in \cite{Biggs2006}.

\begin{rem}
Another way to find 
$B_1,B_2$ is to take into account the symmetry of the directivity:
\begin{equation}
\hat S(\theta,\theta_j) = - \hat S(\theta_j,\theta),
\end{equation}
which is known as the reciprocity theorem. Indeed, then $B_1, B_2$ are found as the solution of the following system of linear equations~\cite{Biggs2006}:
\begin{equation}
\begin{pmatrix}
 \hat S(\theta_1,\theta_1) & \hat S(\theta_1,\theta_2)\\
 \hat S(\theta_2,\theta_1) & \hat S(\theta_2,\theta_2)
\end{pmatrix}
\begin{pmatrix}
 B_1\\
 B_2
\end{pmatrix}
=
\begin{pmatrix}
-\hat S(\theta_i,\theta_1)\\
-\hat S(\theta_i,\theta_2)
\end{pmatrix}.
\end{equation}
\end{rem}

\section{Conclusions}

In this paper we demonstrated how to obtain the embedding formula directly from the Wiener--Hopf formulation where the parameter dependence is given in terms of a pole.   Embedding provides a way of using the Wiener--Hopf problem formulation in cases where a full solution is not known due to the lack of matrix factorization.
The procedure for deriving it is algorithmic and algebraic.
This allows to recover the embedding formulas that are available in the literature. 
We can compare the 
standard procedure (see Introduction) for obtaining the embedding formula with the Wiener--Hopf approach
\begin{enumerate}
  \item After Fourier transform differentiation gets changed into multiplication. So the operator  (\ref{eq: emb_op_strip})
  %\(H=\frac{\partial}{\partial x} +ik \cos \theta_i\) 
  applied to the strip problem correspond to multiplying Wiener--Hopf equation by \(\alpha-ik\cos \theta_i\). 
   \item Likewise for the wedge geometries the operator used is  
   \[
   H=T_p\left( \frac{k}{i}\frac{\partial}{\partial x}\right) -T_p\left(\cos \theta_i\right)
   \]
   and this is linked to the mapping \eqref{eq: var_change} used in deriving the scalar Wiener--Hopf equation.
  \item The homogeneous Wiener--Hopf problem with the correct growth conditions is linked to the edge Green's functions. 
  \item Application of Liouville's theorem provides the uniqueness.
  \item Symmetry of the Wiener--Hopf equation now replaces the reciprocity which is used in obtaining the strong form of embedding. 
  \end{enumerate} 

In deriving the embedding formula the wedge problems were revisited. This was interesting since it is known that the embedding formula for a rational wedge \(q\pi/p\) has \(p\) terms but the known Wiener--Hopf equation for this problem is scalar. The reason for this apparent discrepancy is that the scalar equation is obtained via a mapping and hence does not directly lead to the usual embedding. This has motivated the search for a matrix Wiener--Hopf equation for a wedge which could be obtained with no mappings and a new method of deriving such equations has been presented in Section~\ref{sec: wedge}. Indeed, this new formulation has led to the known embedding formula. Additionally, it is interesting to explore the ways of obtaining embedding formula from the Wiener--Hopf equations derived via mappings and this is described in Section~\ref{map}. This also opens up possibilities to derive more general embeddings since the Wiener--Hopf formulation holds for all angles not just rational multiples of \(\pi\). It is known that the form of embedding formula has to be more complicated due to the lack of suitable local operators to be applied to the total field. The Wiener--Hopf approach described here is a promising avenue to taking more interesting problems of this type. Additionally, other problems which can be reduced to a Wiener--Hopf equation can now be analysed~\cite{HWpaperI, Kisil2023, Medvedeva24, Shanin2022, Makarov2023}.

\vspace*{5mm}
\noindent
{\bf Acknowledgements}
\vspace*{2mm}

\noindent
 A.V.K. is supported by a Royal Society
Dorothy Hodgkin Research Fellowship which also supported A.I.K via the Royal Society Research Fellows Enhanced Research Expenses.
Authors gratefully acknowledge the support of the EU H2020 grant MSCA-RISE-2020-101008140-EffectFact. The authors would also like to thank the Isaac Newton Institute for Mathematical Sciences (INI) for their support and hospitality during the programme ``WHT Follow on: the applications, generalisation and implementation of the Wiener--Hopf Method''(WHTW02), where work on this paper was undertaken and supported by EPSRC grant no EP/R014604/1. The authors are grateful to A.~V.~Shanin for the valuable and insightful discussions.

\bibliographystyle{ieeetr}
\bibliography{Embedding}

\begin{thebibliography}{10}

\bibitem{NSR_20}
D.~Abrahams, X.~Huang, A.~Kisil, G.~Mishuris, M.~Nieves, S.~Rogosin, and I.~Spitkovsky, ``{Reinvigorating the {W}iener--{H}opf technique in the pursuit of understanding processes and materials},'' {\em National Science Review}, vol.~8, 09 2020.
\newblock nwaa225.

\bibitem{Craster2003}
R.~V. Craster, A.~V. Shanin, and E.~M. Doubravsky, ``Embedding formulae in diffraction theory,'' {\em Proceedings of the Royal Society of London. Series A: Mathematical, Physical and Engineering Sciences}, vol.~459, pp.~2475--2496, Oct. 2003.

\bibitem{Sieger1908}
B.~Sieger, ``Die beugung einer ebenen elektrischen welle an einem schirm von elliptischem querschnitt,'' {\em Annalen der Physik}, vol.~332, no.~13, pp.~626--664, 1908.

\bibitem{Colbrook20}
M.~J. Colbrook and A.~V. Kisil, ``A {M}athieu function boundary spectral method for scattering by multiple variable poro-elastic plates, with applications to metamaterials and acoustics,'' {\em Proceedings of the Royal Society A: Mathematical, Physical and Engineering Sciences}, vol.~476, no.~2241, p.~20200184, 2020.

\bibitem{iter_n}
M.~J. Priddin, A.~V. Kisil, and L.~J. Ayton, ``Applying an iterative method numerically to solve n by n matrix {W}iener--{H}opf equations with exponential factors,'' {\em Philosophical Transactions of the Royal Society A: Mathematical, Physical and Engineering Sciences}, vol.~378, no.~2162, p.~20190241, 2020.

\bibitem{Shanin2003}
A.~V. Shanin, ``Diffraction of a plane wave by two ideal strips,'' {\em The Quarterly Journal of Mechanics and Applied Mathematics}, vol.~56, pp.~187--215, May 2003.

\bibitem{tikhonov1948radiation}
A.~N. Tikhonov and A.~A. Samarskii, ``The radiation principle,'' {\em Zhur. Eksptl. i Teoret. Fiz.}, vol.~18, 1948.

\bibitem{Nazarov1994}
S.~A. Nazarov and B.~A. Plamenevsky, {\em Elliptic Problems in Domains with Piecewise Smooth Boundaries}.
\newblock DE GRUYTER, Dec. 1994.

\bibitem{Babic2008}
V.~M. Babič, M.~A. Lyalinov, and V.~E. Grikurov, {\em Diffraction theory}.
\newblock Alpha Science series on wave phenomena, Oxford: Alpha Science, 2008.

\bibitem{Gibbs_2018}
A.~Gibbs, S.~Langdon, and A.~Moiola, ``Numerically stable computation of embedding formulae for scattering by polygons,'' 2018.

\bibitem{WILLIAMS1982}
M.~H. Williams, ``Diffraction by a finite strip,'' {\em The Quarterly Journal of Mechanics and Applied Mathematics}, vol.~35, no.~1, pp.~103--124, 1982.

\bibitem{porter1991solution}
D.~Porter, ``The solution of integral equations with difference kernels,'' {\em The Journal of Integral Equations and Applications}, pp.~429--454, 1991.

\bibitem{Biggs2006}
N.~R.~T. Biggs, ``A new family of embedding formulae for diffraction by wedges and polygons,'' {\em Wave Motion}, vol.~43, pp.~517--528, Aug. 2006.

\bibitem{Noble1958}
B.~Noble, {\em Methods Based on the {W}iener-{H}opf Technique for the Solution of Partial Differential Equations}.
\newblock International Series of Monographs on Pure and Applied Mathematics. Vol. 7, New York: Pergamon Press, 1958.

\bibitem{myWHreview}
A.~V. Kisil, I.~D. Abrahams, G.~Mishuris, and S.~V. Rogosin, ``The {W}iener-{H}opf technique, its generalizations and applications: constructive and approximate methods,'' {\em Proc. A.}, vol.~477, no.~2254, pp.~Paper No. 20210533, 32, 2021.

\bibitem{gakhov1952riemann}
F.~D. Gakhov, ``Riemann's boundary problem for a system of n pairs of functions,'' {\em Uspekhi matematicheskikh nauk}, vol.~7, no.~4, pp.~3--54, 1952.

\bibitem{Muskhelishvili1977}
N.~I. Muskhelishvili, {\em Singular Integral Equations}.
\newblock Springer Netherlands, 1977.

\bibitem{Dan_Lomb07}
V.~Daniele and G.~Lombardi, ``Fredholm factorization of {W}iener--{H}opf scalar and matrix kernels,'' {\em Radio Science}, vol.~42, no.~6, 2007.

\bibitem{Olver2014}
F.~W.~J. Olver, {\em Asymptotics and Special Functions}.
\newblock Burlington: Elsevier Science, 2014.
\newblock Description based upon print version of record.

\bibitem{Shanin2015}
A.~V. Shanin and A.~I. Korolkov, ``Diffraction by an impedance strip {I}. {R}educing diffraction problem to {R}iemann–{H}ilbert problems,'' {\em The Quarterly Journal of Mechanics and Applied Mathematics}, vol.~68, pp.~321--339, July 2015.

\bibitem{shanin1998excitation}
A.~V. Shanin, ``Excitation of waves in a wedge-shaped region,'' {\em Acoustical Physics}, vol.~44, no.~5, pp.~592--597, 1998.

\bibitem{Aitken2024}
M.~Aitken, {\em On the factorisation of matrix Wiener--Hopf kernels arising from acoustic scattering problems}.
\newblock PhD thesis, Apollo - University of Cambridge Repository, 2024.

\bibitem{Sommerfeld1896}
A.~Sommerfeld, ``Mathematische theorie der diffraction: Mit einer tafel,'' {\em Mathematische Annalen}, vol.~47, pp.~317--374, June 1896.

\bibitem{malyuzhinets1955radiation}
D.~Malyuzhinets, ``Radiation of sound from the vibrating faces of an arbitrary wedge [part {II}],'' {\em Sov. Phys. Acoust}, vol.~1, pp.~240--248, 1955.

\bibitem{Lyalinov2005}
M.~A. Lyalinov, ``Generalized {S}ommerfeld integral and diffraction in an angle-shaped domain with a radial perturbation,'' {\em Journal of Physics A: Mathematical and General}, vol.~38, pp.~L707--L714, Oct. 2005.

\bibitem{shanin2020sommerfeld}
A.~V. Shanin and A.~I. Korolkov, ``{S}ommerfeld-type integrals for discrete diffraction problems,'' {\em Wave Motion}, vol.~97, p.~102606, 2020.

\bibitem{Shanin2022}
A.~V. Shanin and A.~I. Korolkov, ``{D}iffraction by a {D}irichlet right angle on a discrete planar lattice,'' {\em Quarterly of Applied Mathematics}, vol.~80, pp.~277--315, Feb. 2022.

\bibitem{Craster2005}
R.~V. Craster and A.~V. Shanin, ``Embedding formulae for diffraction by rational wedge and angular geometries,'' {\em Proceedings of the Royal Society A: Mathematical, Physical and Engineering Sciences}, vol.~461, pp.~2227--2242, June 2005.

\bibitem{Daniele2020}
V.~Daniele and G.~Lombardi, ``The {W}iener--{H}opf theory for the scattering by an impenetrable polygonal structure,'' in {\em 2020 IEEE International Symposium on Antennas and Propagation and North American Radio Science Meeting}, pp.~723--724, IEEE, July 2020.

\bibitem{Daniele_wedgebook20}
V.~Daniele and G.~Lombardi, {\em Scattering and Diffraction by Wedges {1}: The {W}iener--{H}opf Solution – Theory}.
\newblock John Wiley \& Sons, Ltd, 09 2020.

\bibitem{Shanin2010}
A.~V. Shanin and R.~V. Craster, ``Pseudo-differential operators for embedding formulae,'' {\em Journal of Computational and Applied Mathematics}, vol.~234, pp.~1637--1646, July 2010.

\bibitem{Nethercote2020}
M.~Nethercote, R.~Assier, and I.~Abrahams, ``Analytical methods for perfect wedge diffraction: A review,'' {\em Wave Motion}, vol.~93, p.~102479, Mar. 2020.

\bibitem{HWpaperI}
M.~A. Nethercote, A.~V. Kisil, and R.~C. Assier, ``{Diffraction of acoustic waves by a wedge of point scatterers},'' {\em SIAM J. Appl. Math.}, vol.~82, no.~3, pp.~872--898, 2022.

\bibitem{Kisil2023}
A.~V. Kisil, ``{A generalisation of the Wiener--Hopf methods for an equation in two variables with three unknown functions.},'' {\em {SIAM J. Appl. Math.}}, vol.~84, no.~2, pp.~464--476, 2024.

\bibitem{Medvedeva24}
E.~Medvedeva, R.~Assier, and A.~Kisil, ``Diffraction by a set of collinear cracks on a square lattice: An iterative {W}iener–-{H}opf method,'' {\em WAVE MOTION}, Apr. 2024.

\bibitem{Makarov2023}
O.~I. Makarov, A.~V. Shanin, and A.~I. Korolkov, ``The {S}ommerfeld integral in problems of simulating the diffraction of acoustic waves using a triangular lattice,'' {\em Acoustical Physics}, vol.~69, pp.~143--158, Apr. 2023.

\bibitem{Daniele1984}
V.~G. Daniele, ``On the solution of two coupled {W}iener–-{H}opf equations,'' {\em SIAM Journal on Applied Mathematics}, vol.~44, pp.~667--680, Aug. 1984.

\bibitem{Khrapkov1971}
A.~A. Khrapkov, ``The first basic problem for a notch at the apex of an infinite wedge,'' {\em International Journal of Fracture Mechanics}, vol.~7, pp.~373--382, Dec. 1971.

\end{thebibliography}

\appendix
\appendixpage
\section{Wiener--Hopf equation}
\label{app: W-H}

For completeness we are going to outline the usual
procedure used to solve the Wiener--Hopf problem~\cite{Noble1958}. Although it is not necessary~\cite{myWHreview} the region of analyticity will be extended, in order to introduce a strip of overlap between the upper and lower half-planes. Define {\it upper half-plane} ${\mathcal H}^{+} =
\{{\mathrm{Im}}\, z > a; a < 0\}$, {\it lower half-plane} ${\mathcal H}^{-} =
\{{\mathrm{Im}}\, z < b; b > 0\}$ and {\it the Wiener--Hopf strip} ${\mathcal H} =
{\mathcal H}^{+}\cap {\mathcal H}^{-} = \{a < {\mathrm{Im}}\, z < b\}$
($a$ and $b$ are real numbers, typically small). The superscript of \(\pm\) will indicate the region of analyticity. 
So our equation is (\ref{eq: matrix_WH}).
% \begin{equation}
% \label{eq: matrix_WH_app}
% \rmU^-(z,z_i)  = {\rm K}(z) {\rmU}^+(z,z_i) + \rmF(z,z_i),\quad z \in     {\mathcal H},
% \end{equation}
%where \(z_i\) is a parameter.
The key step in the solution is to construct a factorisation of the kernel \({\rm K}(z)\) (this step is straightforward for scalar kernels but in general no procedure exists for matrix kernels)
\begin{equation}
\label{eq: factors_WH}
{\rmA}(z)={\rmA^-}(z){\rmA^+}(z),
\end{equation}
where the factors \({\rmA^-}(z)\) and their inverses are analytic in  \({\mathcal H^{\pm}}\).
Multiplying \eqref{eq: matrix_WH} through by \((\rmA^-)^{-1}\) we obtain:
\begin{equation}
\label{eq: matrix_WH1}
{(\rmA^-(z))^{-1}}  \rmU^-(z,z_i)  = {\rmA^+}(z) {\rmU}^+(z,z_i) + {(\rmA^-(z))^{-1}} \rmF(z,z_i),
\end{equation}
Now performing an \emph{additive splitting} we have (this step is straightforward for both matrix and scalar functions):
\[ {(\rmA^-(z))^{-1}} \rmF(z,)=[{(\rmA^-(z))^{-1}} \rmF(z,z_i)]^-+[{(\rmA^-(z))^{-1}} \rmF(z,z_i)]^+,\quad z\in {\mathcal H},\]
where \(  [{(\rmA^-(z))^{-1}} \rmF(z,z_i)]^{\pm}\) are analytic in their indicated half-planes, respectively.
Finally rearranging the equation yields
\[
{(\rmA^-(z))^{-1}}  \rmU^-(z,z_i) - [{(\rmA^-(z))^{-1}} \rmF(z,z_i)]^-=
\]
\begin{equation}
\label{WH_fact_estimates}
 {\rmA^+}(z) {\rmU}^+(z,z_i) + [{(\rmA^-(z))^{-1}} \rmF(z,z_i)]^+,
\end{equation}
The right hand side of \eqref{WH_fact_estimates} is a function
analytic in ${\mathcal H}^{+}$ and the left hand side of
\eqref{WH_fact_estimates} is analytic in ${\mathcal H}^{-}$. Hence, together they offer
analytic continuation from ${\mathcal H}$ into the whole  $z$-plane, and so
each side is equal to an entire function $J(z)$. We also require \emph{at most
algebraic growth} of each side of \eqref{WH_fact_estimates} as $|z| \to \infty$ in the
respective half-planes of analyticity, i.e.\
\[J(z)\leq {\cal O} (|z|^{n}), \quad |z| \to \infty,\]
for some constant $n$. Applying the extended form of Liouville's theorem implies that
\(J(z)\) is a polynomial of degree less than or equal to
\(n\). Hence \(\rmU^-(z,z_i)\) and \({\rmU}^+(z,z_i)\) are
determined up to \(\lfloor n \rfloor +1\) unknown constants (typically \(n\leq 0\) or
\(1\), and the constants can be fixed using extra information about the behaviour of the solution).

\section{Alternative derivation of plane wave embedding}
\label{app: alter_emb}

Consider a set of $N$ solutions ${\rmU(z,z_l)}$ of (\ref{eq: matrix_WH}) with pole forcing (\ref{eq:pole_forcing}) for different values of parameter $z_l$ and index them as  $(l= 1, \dots,N)$.  The set is not normal, since solutions have poles that correspond to the forcing. Let us modify this  set as follows: 
\[
\hat \rmU(z,z_l) = (z-z_l)\rmU(z,z_l).
\]
The matrix 
\[
\rmX^\pm_p(z) = \left(\hat \rmU(z,z_1),\ldots,\hat \rmU(z,z_N)\right)
\]
composed of modified solutions satisfy the following equation: 
\[
\rmX^-_p(z)  - {\rm R}= \rmA\rmX^+_p(z), ,\quad {\rm R} = ({\rm r},\ldots,{\rm r})
\]
Assuming that ${\rm det}(\rmX^+_p)$ is non-zero we can take 
$
\rmX^-_p(z)  - {\rm R}
$
as the matrix of normal solutions in the lower half-plane, and 
$\rmX^+_p(z)$
as the matrix of normal solutions in the upper half-plane. Then, using the canonical embedding formulae (\ref{eq:Can_solution_pole2}) we obtain:

\begin{equation}
\label{eq:plane_solution_pole2}
{\rmU}^{+}(z,z_i) = {\rmX}_p^{+}(z){\rm P}(z)-\frac{{\rm X}_p^{+}(z)\left(\rmX^-_p(z_i)  - {\rm R}\right)^{-1}{\rm r}}{z-z_i}. 
\end{equation}

\begin{rem}
If the set of parameters $z_l$ does not contain two or more equal values, then the determinant of matrices ${\rm det}(\rmX^+_p)$ can turn into zero only in a finite set of points. It was shown in \cite{Muskhelishvili1977} that by simple algebraic manipulations, these matrices can be transformed into normal ones. However, it appears that for diffraction problems, at least for the ones considered in this paper ${\rm det}(\rmX^\pm_p)$ is non-zero everywhere, and no additional modifications are needed.   
\end{rem}

\section{Wiener--Hopf equation for the half-plane}
\label{app: halfline}

\begin{figure}[ht]
	\centering{
	\includegraphics[width=0.5\textwidth]{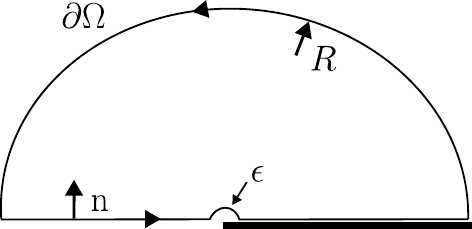}
	}
	\caption{Integration contours for Green's second identity}
	\label{fig:Greens_contour_halfplane}
\end{figure}

Below we give a derivation of the Wiener--Hopf problem for the problem of diffraction by a half-plane.
First, let us split the total field into symmetrical and antisymmetrical parts:
\begin{equation}
\label{eq: field_sym}
u = u^{\rm a}(x,y) + u^{\rm s}(x,y) = 1/2(u(x,y) - u(x,-y)) + 1/2(u(x,y) +u(x,-y)). 
\end{equation}
Notice that the antisymmetrical part of the scattered field is equal to zero, and the consideration can be restricted only to the upper half with the mixed boundary conditions on line $x=0$:
\[
\frac{\ptl u^{\rm sc}}{\ptl y}(x,0) = 0, \quad x<0 \quad u^{\rm sc}(x,0) = - \exp\{-ikx\cos\theta_i\},\quad \quad x>0
\]
Introduce an outgoing wave
\[
w = \exp\{iz x  + i\gamma(z)y\},\quad \gamma(z) = \sqrt{k^2 - z^2}
\]
Apply  Green's second identity
\begin{equation}
\label{eq: Greens_theorem}
\int_{\ptl \Omega }\left[w\frac{\ptl u^{\rm sc}}{\ptl n}-u^{\rm sc}\frac{\ptl w}{\ptl n}\right]dl=0,
\end{equation}
where the contour of integration is shown in Figure~\ref{fig:Greens_contour_halfplane}.
After taking the limits $\varepsilon \to 0$ and $R\to \infty$ obtain the following functional relation
\begin{equation}
\label{Wh_half_plane}
U^-(z) - \frac{1}{i\gamma(z)} U^+ - \frac{i}{z - \kstar} = 0, \quad \kstar = k\cos\theta_i
\end{equation}
where
\begin{equation}
\label{eq:a_coef_2}
U^-(z)  = \int_{-\infty}^{0} \exp\{iz x\}u^{\rm sc}(x,0) dx,\quad U^+(z)  = \int_{0}^{\infty} \exp\{iz x\}\frac{\ptl u^{\rm sc}}{\ptl y}(x,0) dx.
\end{equation}
The latter defines function $U^-(z)$ and $U^+(z)$ analytical in lower and upper half-planes correspondingly. Thus (\ref{Wh_half_plane}) is a scalar Wiener--Hopf problem on the line $(-\infty,\infty)$. 
The behaviour of the solution at infinity is defined by Meixner's condition. Indeed, it follows from (\ref{Meixner_cond}) and Watson's lemma that 
\[
U^{-}(z) \sim {\frac{C_1}{z^{3/2}}}, 
\quad
U^{+}(z) \sim \frac{C_2}{\sqrt{z}}, 
\]
as $z \to \infty$ in the lower and the upper half-planes correspondingly, and $C_1, C_2$ are some unknown constants. Thus $U^-$ and $U^+$ tend to zero at infinity.

Let us consider the problem for the edge Green's function (\ref{eq:edge_greens_plane}). Careful study \cite{Shanin2003} shows that the edge Green's function behaves as follows near the edge of the half-plane:
\[
v(r,\theta) = -\frac{1}{\sqrt{\pi}}r^{-1/2}\sin(\theta/2) + O(r^{3/2}).
\]
Thus, Fourier integrals with the normal derivative diverge in the vicinity of the edge. However, the integrals can be regularised by deforming the contour of integration in the vicinity of the edge and then studying the limit. Indeed, let us apply Green's second identity with $v$ and $w$ and the contour of integration shown in Figure~\ref{fig:Greens_contour_halfplane}, to obtain
\[
V^-(z) - \frac{1}{i\gamma(z)}V^+(z)=0,
\]
where 
\[
V^-(z)  = \int_{-\infty}^{0} \exp\{iz x\}v(x,0) dx,\quad V^+(z)  = \lim_{\epsilon\to 0}\int_\Gamma \exp\{iz x\}\frac{\ptl v}{\ptl y}(x,0) dx.
\]
where $\Gamma$ is union of a half-line $(\epsilon,\infty)$ and a circle of radius $\epsilon$ having center in the origin. The limit as $\epsilon\to 0$ provides the regularization of the improper integral for $V^+(z)$. Also, applying Watson's lemma we get the following growth estimations:
\[
V^{-}(z) \sim \sqrt{\frac{i}{z}},\quad
V^{+}(z) \sim i\sqrt{iz} 
\]
as $z \to \infty$ in the lower and the upper half-planes correspondingly.

Note that following a similar procedure one can derive functional equations for edge Green's functions of an arbitrary number of co-linear strips.
\section{Wiener--Hopf equation for the strip}
\label{app: strip}
Consider the problem of diffraction by a finite strip introduced in the introduction. Splitting the problem into symmetrical and antisymmetrical parts (see~(\ref{eq: field_sym})) the problem is restricted to the upper half-plane with the following mixed boundary conditions:
\[
\frac{\ptl u^{\rm sc}}{\ptl y}(x,0) = 0, \quad x \in [-\infty,a)\&(a,\infty],
\]
\[
u^{\rm sc}(x,0) = - \exp\{-ikx\cos\theta_i\},\quad  x \in [-a,a].
\]
% Let the total field be presented as a sum of scattered and diffracted fields (\ref{eq:inc_field}). Let the scattered field satisfy radiation conditions, introduce the directivity of the scattered field similar to (\ref{eq:Direct_sc}), and impose Meixner conditions (\ref{Meixner_cond}) in each vertex of the strip ($r$ and $\theta$ in (\ref{Meixner_cond}) is treated as local coordinates near the edge).
%\subsubsection{Derivation of  Matrix Wiener Hopf equations}
Apply Green's theorem in the upper half-plane (see also the domain for the half-plane problem~\ref{fig:Greens_contour_halfplane}). Obtain the following functional equation:
\begin{equation}
\label{eq:func_strip}
U^-(z) - \frac{1}{i\gamma(z)}U^0(z)+U^+(z) = \frac{ie^{-i(z-\kstar)a}}{z-\kstar},
\end{equation} 
where
\[
U^-(z)  = \int_{-\infty}^{-a} \exp\{iz x\}u^{\rm sc}(x,0) dx ,\quad U^0(z)  = \int_{-a}^{a} \exp\{iz x\}\frac{\ptl u^{\rm sc}}{\ptl y}(x,0) dx,
\]
\[
U^+(z) = \int_a^\infty \exp\{iz x\}u^{\rm sc}(x,0)dx + \frac{ie^{i(z-\kstar)a}}{z-\kstar}.
\]
 It follows from the latter that $U^{-}(z)$ is analytical in the lower half-plane, $U^{+}(z)$ analytical in the upper half-plane, and $U^{0}(z)$ is an entire function. Equation (\ref{eq:func_strip}) was extensively studied in (\cite{Shanin2003,Shanin2015}), and it is not of Wiener--Hopf type, since it involves tree unknown functions. Another difficulty is that unknown functions grow exponentially at infinity. Indeed, it follows from Meixner conditions that:
\[
U^{-}(z) \sim -{\frac{\sqrt{i}e^{-iz a}}{z^{3/2}}}, \quad
U^{0}(z) \sim -\frac{i^{3/2}e^{iz a}}{\sqrt{z}}, 
\]
in the lower half-plane as $z \to \infty$,
\[
U^{0}(z) \sim -\frac{i^{3/2}e^{-iz a}}{\sqrt{z}}, \quad
U^{+}(z) \sim {\frac{ie^{i(z-z_i) a}}{z}}, 
\]
in the upper half-plane as $z \to \infty$.

Let us reformulate (\ref{eq:func_strip}) as a matrix Wiener--Hopf problem. Denote:
\[
 U^-_1 = e^{i z a}U^{-}(z), \quad  U^+_1 = e^{-i z a}U^{+}(z),\quad U^+_2 = e^{iz a}U^0(z),\quad U^-_2 = e^{-iz a}U^0(z).
\]
Functions $U^-_i$ are analytical in the lower half-plane and have power growth there, while $U^+_{i}$ are analytical in the upper half-plane and have power growth there. Thus, we have (\ref{eq: matrix_WH}) with
\[
{\rm U^\pm}=
\begin{pmatrix}
U^\pm_1 \\
U^\pm_2
\end{pmatrix},
\quad
{\rm A}=
\begin{pmatrix}
-\exp\{2iz a\} &  (i\gamma(z))^{-1} \\
0 &\exp\{-2iz a\}
\end{pmatrix},\quad
{\rm F} = \frac{i \exp\{i\kstar a\}}{z-\kstar}
\begin{pmatrix}
1\\
0
\end{pmatrix}.
\]
\section{Explicit solution to the right-angled wedge problem}
\label{app:wedge}
Let's factorize $\rmA$ from (\ref{eq: WH_edge_final}), by reducing it to the matrix of Daniele-Khrapkov type:
\begin{equation}
\begin{pmatrix}
1& 0\\
0& -1
\end{pmatrix}{\rmA} =
\frac{1}{2\gamma}
\begin{pmatrix}
1& -3i\gamma^2\\
-i& 1
\end{pmatrix}.
\end{equation}  
Following \cite{Daniele1984,Khrapkov1971}, one then can factorize the kernel:
\begin{equation}
{\rmA}(z)=
\begin{pmatrix}
-a(z)&-3\gamma^2b(z)\\
b(z)&a(z)
\end{pmatrix}
\begin{pmatrix}
a(-z)&3\gamma^2b(-z)\\
b(-z)&a(-z)
\end{pmatrix},  
\end{equation}
where 
\[
a(z) = \cos\left(\frac{1}{3}\theta(z)\right), \quad b(z) = -\frac{i}{\sqrt{3}\gamma(z)}\sin\left(\frac{1}{3}\theta(z)\right),\quad
\theta(z) = \arccos\left(\frac{z}{k}\right),
\]
and matrices on the right-hand side are analytical in the upper and the lower half-planes correspondingly. Columns of these matrices can be considered as solutions of the homogeneous Wiener--Hopf problem. However, they do not satisfy growth conditions (\ref{growth_cond_can_wedge}) which correspond to the edge Green's function. Studying the linear combinations of these solutions with polynomial coefficients one can find the solutions satisfying the growth conditions, and build the matrix $\tilde \rmX$:
\begin{equation}
\renewcommand{\arraystretch}{2}
{\tilde\rmX}^- (z) =
\begin{pmatrix}
\tilde C_{11}\cos\left(\dfrac{2}{3}\theta(-z)\right) & \tilde C_{12}\cos\left(\dfrac{4}{3}\theta(-z) \right) \\
\tilde C_{21}\gamma^{-1}(z)\sin\left(\dfrac{2}{3}\theta(-z)\right)  & \tilde C_{22}\gamma^{-1}(z)\sin\left(\dfrac{4}{3}\theta(-z)\right) 
\end{pmatrix},
\label{Canon_wedge_exp}
\end{equation}
where
\[
\tilde C_{11} = -2\left(-\dfrac{k}{2}\right)^{\dfrac{2}{3}}\Gamma\left(-\frac{2}{3}\right),\quad \tilde C_{12} = 4\left(-\dfrac{k}{2}\right)^{\dfrac{4}{3}}\Gamma\left(-\frac{4}{3}\right),
\]
\[
\tilde C_{21} = -i\sqrt{3}\left(-\dfrac{k}{2}\right)^{\dfrac{2}{3}}\Gamma\left(\frac{1}{3}\right),\quad \tilde C_{22} = -i\sqrt{3}\left(-\dfrac{k}{2}\right)^{\dfrac{4}{3}}\Gamma\left(-\frac{1}{3}\right).
\]
% Detailed study shows that to find the solution of the original diffraction problem it is enough to find $U_1(z)$.  Instead of doing it, let us find the directivity of the scattered field. Let, for definiteness, the angle of observation $\theta$ be restricted in the upper half-plane $0<\theta<\pi$. Then, applying Green's theorem in the upper half-plane to the scattered field $u^{\rm sc}$ and
% \[
% w = \exp\{-i kx\cos\theta - iky\sin\theta \} - \exp\{-ikxk\cos\theta + iky\sin\theta\},
% \]   
% after a bit of algebra, we obtain the following expression:
Thus, from (\ref{Wedge_sol}) we obtain the following expression for the directivity of the scattered field:
% \begin{equation}
% S(\theta,\theta_i) = -k\sin\theta_iU_1(-k\cos\theta).
% \end{equation}
% Let us check that it is consistent with the known solution \cite{Craster2003}. Indeed, using (\ref{Wedge_sol}) and (\ref{Canon_wedge_exp}) we get

\begin{equation}
S(\theta,\theta_i) = -\dfrac{2i}{\sqrt{3}}\dfrac{\sin\left(\dfrac{4}{3}\theta_i\right)\sin\left(\dfrac{2}{3}\theta \right) -\sin\left(\dfrac{2}{3}\theta_i \right)\sin\left(\dfrac{4}{3}\theta \right)}{\cos^2\theta - \cos^2\theta_i},
\end{equation} 
which is consistent with (3.16)  of \cite{Craster2005}.

\end{document}